# Electromagnetics with time-varied vacuum permittivity and permeability

ZHIWEI SUN,[1]

[1] *School of Microelectronics and Communication Engineering, Chongqing University, Chongqing, 400044, China*

**Abstract:** Maxwell equations provide a complete description of the electromagnetic (EM) phenomena, which have been one of the key fundamental-theories of modern physics, such as electromagnetism, optics, quantum theories, etc. The vacuum permittivity and permeability (P&P) in the constitutive relation were regarded as constant, resulting into that the vacuum lightspeed is constant with time. However, neither the Maxwell equations nor any experiments demonstrated that the P&P must be invariable with time. Here, the P&P are assumed as time-varied, achieving a mathematical result exhibits that the P&P are increased with time exponentially. Then, the Maxwell equations of static and covariant EM fields are presented, which include the different constitutive relations. The EM characters in the time-varied case are investigated that, the lightspeed, level EM intensity, and magnetic effect of charge-motion decreases with time. The EM energy transforms into a new energy form caused by the P&P variation, and a new momentum-tensor form caused by propagation is found as well. At last, the Maxwell equations based on the observation results at our time are presented. Meanwhile, since the level EM intensity decreases with time, the energy per frequency is time-decreased, resulting into that the Planck constant decreases exponentially, and correspondingly, the Rydberg constant increases with time. A red-shift of absorption spectrum is predicted based on this finding, which is coincident with the Hubble's observation results.

## 1. Introduction

Maxwell equations [1], including the four divergence- / curl-equations, electric-current definition, and constitutive relation, describe the electromagnetic (EM) phenomena completely. The results on and the equations have been the key foundation of modern physics. In the constitutive relation, the vacuum permittivity and permeability (P&P), which denote, respectively, the electric and magnetic field-to-induction relationships, were generally regarded as constants with time. This develops a result that the vacuum lightspeed is a constant, which has been the foundation of relativity theory and quantum research, etc. However, neither the Maxwell equations nor any experiments limit that the P&P must be invariable with time. Such as, the Michelson-Morley's experiment proved the lightspeed is independent on inertial systems, but it did not demonstrate the lightspeed must be time-independent. Therefore, there exists an opportunity to discuss the EM phenomena with time-varied P&P. In this study, the P&P are assumed time-varied firstly. The mathematical results based on five rational assumptions show that the P&P vary with time exponentially. Then, the Maxwell equations in the time-varied P&P case are obtained, that the static and covariant EM fields are regularized by the equation-terms with different constitutive relations. The EM characters of propagation, force, energy and momentum are investigated. The wavelength, propagation velocity, level EM magnitude, EM force between charges (currents), and force of EM wave are decreased with time. The EM energy decreases with time, since it transforms into a new energy form caused by the time-varied P&P. The EM momentum remains constant over time, however, a new form of momentum density tensor, i.e., a new EM force form, is found in the propagation EM state. At last, the Maxwell equations is defined based on the observation results at our time. The theoretical results predicted based on this finding are compared with the Hubble's observation results [2]. In the universe with time-varied P&P, the observed results resemble being immersed in the "water" with lower P&P, that first there exists an optical refraction in the observed image, and second a red-shift of the absorption spectrum is presented owing to the decrease of Planck constant.

## 2. Assumption and research model

Five assumptions are introduced firstly as the foundation of this research:

**I.** The vacuum P&P vary with time and, this variation is position-independent, since the universe is roughly smooth, i.e., there exists no special position in the universe.

**II.** Maxwell equations, which have been mathematically and experimentally validated thoroughly, can describe the EM phenomena accurately in the small time and space scales. Therefore, the physical results of and the equations can be used as the fundament results of this research and, the new equations should be identical to the classical Maxwell equations when the time and space are limited to small scales.

**III.** Electric and magnetic waves are covariant, i.e., they have the identical propagation feature in vacuum, since, first, they are induced by each other mutually, second, they have the duality relation, and third, the alternating electric and magnetic fields contribute, mathematically, the equal role in the EM propagation.

**IV.** EM energy can transfer from the EM fields to space only, i.e., EM field cannot obtain energy from the free space, according to the law of conservation of energy and the entropy increase principle.

**V.** Static EM field and EM waves have the same propagation velocity, being equivalent to the lightspeed that time.

Note that, there exist four supporting assumptions II to V, however, they are completely coincident with the classical research. So that, the different EM phenomena exhibited in this study are completely caused by the assumption I that the P&P are time-varied.

The research model is set as follows for better introducing this study:

**I.** The research is studied in free space, i.e., there exist no EM polarity-effect induced by the polar molecule. So that, the P&P mentioned in this research are the vacuum P&P, which are scalar quantities.

**II.** The relativity effect is not considered in this research, since our calculation does not introduce any moving inertial systems or gravitation.

**III.** Maxwell equations are independent on the coordinate systems selected. Herein, the Cartesian coordinate system is employed to present the propagation features as we focus on the physical discussion rather than the mathematical solution of Maxwell equations in different coordinates.

## 3. Variation function of permittivity and permeability

Based on the assumption I, the vacuum permittivity (denoted by $\varepsilon_0$) and permeability (denoted by $\mu_0$) are time-varied, which can be written as $\varepsilon_0(t)$ and $\mu_0(t)$, respectively. The covariant EM field in source-free case is used to define the variation function of P&P, of which the schematics of equation $\nabla\times\mathbf{E}=-\partial\mathbf{B}/\partial t$ and $\nabla\times\mathbf{H}=\partial\mathbf{D}/\partial t$ are shown in Figs. 1(a) and (b), respectively. The investigation region can be confined in a small space-scale since: (i) the electric and magnetic induce each other at the same time and in same position, and (ii) the P&P are position-independent. In this case, only the variation-with-time of P&P is considered, since the propagation feature can be neglected in the confined space.

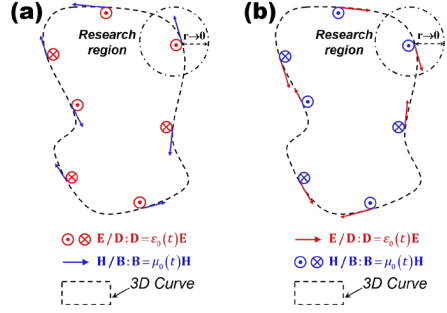

Fig. 1. Covariant EM field in confined region. (a) Schematic of equation $\nabla \times \mathbf{E} = -\partial \mathbf{B}/\partial t$. (b) Schematic of equation $\nabla \times \mathbf{H} = \partial \mathbf{D}/\partial t$. The EM fields are investigated in a confined region (the radius of the research region $r$ tends to zero), so that the classical Maxwell equations can be used to present the covariant EM waves by only introducing the derivative of P&P to time.

Based on the assumption II, the classical Maxwell equations, in which the P&P are defined to be time-varied, is used to describe the covariant EM field here, which is given as:

$$\begin{cases} \nabla \times \mathbf{H} = \varepsilon_0 \dfrac{\partial \mathbf{E}}{\partial t} + \mathbf{E}\dfrac{\partial \varepsilon_0}{\partial t} \\ \nabla \times \mathbf{E} = -\mu_0 \dfrac{\partial \mathbf{H}}{\partial t} - \mathbf{H}\dfrac{\partial \mu_0}{\partial t} \\ \nabla \cdot \mathbf{B} = 0 \\ \nabla \cdot \mathbf{D} = 0 \\ \textbf{Constitutive relation: } \mathbf{D} = \varepsilon_0 \mathbf{E} \quad \mathbf{B} = \mu_0 \mathbf{H} \\ \qquad \varepsilon_0 = \varepsilon_0(t) \quad \mu_0 = \mu_0(t) \end{cases} \quad , \qquad (1)$$

where, $\mathbf{E}$ and $\mathbf{H}$ are the electric and magnetic field intensity, respectively; $\mathbf{D}$ and $\mathbf{B}$ are the electric and magnetic induction intensity, respectively. Obviously, Eq. (1) will transform into the classical equations if let the P&P be time-independent. Construct equations of $\nabla \times (\nabla \times \mathbf{H})$ and $\nabla \times (\nabla \times \mathbf{E})$ as:

$$\begin{cases} -\dfrac{\partial \varepsilon_0}{\partial t}\left(\dfrac{\partial \mu_0}{\partial t}\mathbf{H} + \mu_0 \dfrac{\partial \mathbf{H}}{\partial t}\right) - \varepsilon_0\left(\dfrac{\partial^2 \mu_0}{\partial t^2}\mathbf{H} + 2\dfrac{\partial \mu_0}{\partial t}\dfrac{\partial \mathbf{H}}{\partial t} + \mu_0 \dfrac{\partial^2 \mathbf{H}}{\partial t^2}\right) + \nabla^2 \mathbf{H} = 0 \\ -\dfrac{\partial \mu_0}{\partial t}\left(\dfrac{\partial \varepsilon_0}{\partial t}\mathbf{E} + \varepsilon_0 \dfrac{\partial \mathbf{E}}{\partial t}\right) - \mu_0\left(\dfrac{\partial^2 \varepsilon_0}{\partial t^2}\mathbf{E} + 2\dfrac{\partial \varepsilon_0}{\partial t}\dfrac{\partial \mathbf{E}}{\partial t} + \varepsilon_0 \dfrac{\partial^2 \mathbf{E}}{\partial t^2}\right) + \nabla^2 \mathbf{E} = 0 \end{cases} . \quad (2)$$

Based on the assumption III, $\mathbf{E}$ and $\mathbf{H}$ are covariant which have the same propagation equations, so that:

$$\begin{cases} \varepsilon_0 \dfrac{\partial \mu_0}{\partial t} = \mu_0 \dfrac{\partial \varepsilon_0}{\partial t} \\ \varepsilon_0 \dfrac{\partial^2 \mu_0}{\partial t^2} = \mu_0 \dfrac{\partial^2 \varepsilon_0}{\partial t^2} \end{cases} . \qquad (3)$$

The mathematical solutions of Eq. (3) demonstrate that both $\varepsilon_0$ and $\mu_0$ vary with time exponentially, which is formulated as (detailed in Supplementary material S1):

$$\begin{cases} \varepsilon_0 = \varepsilon_0^o e^{At} \\ \mu_0 = \mu_0^o e^{At} \end{cases}, \tag{4}$$

in which, $\varepsilon_0^o$ and $\mu_0^o$ signify the original values of $\varepsilon_0$ and $\mu_0$, and $A$ is the variation gradient of the exponential term of $\varepsilon_0$ and $\mu_0$. Note that, the $\varepsilon_0^o$ and $\mu_0^o$, as well as the $\varepsilon_0$ and $\mu_0$, here are introduced to present a general relationship of **E-D** and **H-B**, respectively. They are the relationship factors, which differ from the classical P&P with the certain values based on measurements.

The sign of $A$ can be defined by analyzing the EM energy flux density (detailed in Supplementary material S2), that the Poynting's theorem is given as:

$$\nabla \cdot (\mathbf{E} \times \mathbf{H}) = -\frac{1}{2}\left(\varepsilon_0 \frac{\partial}{\partial t} E^2 + \mu_0 \frac{\partial}{\partial t} H^2\right) - \left(E^2 \frac{\partial \varepsilon_0}{\partial t} + H^2 \frac{\partial \mu_0}{\partial t}\right). \tag{5}$$

In vacuum, the energy-flow included in the left term equal to zero. The first term on the right, which signifies the EM stored-energy, should be positive, since the energy can only transfer from EM fields to space based on the assumption IV. So that, the second term should be negative that:

$$-A\left(\varepsilon_0 E^2 + \mu_0 H^2\right) < 0. \tag{6}$$

Obviously, $A$ is positive, that P&P are increased with time following the formula $\exp(At)$. Evidently, the value of $A$ must be extremely low since no observed result indicates an obvious variation of P&P.

### 4. Electromagnetics with time-varied vacuum permittivity and permeability

#### 4.1 Electromagnetics in confined region

The electromagnetics in the confined region is discussed in this section. The covariant EM field has been presented in Eq. (1), so we focused on the static electric / magnetic field here. Historically, the field-effect of charge and current was defined by **E** and **B**, respectively. Nonetheless, the **E-H** and **D-B** relationships are more appreciated in the mathematical system of Maxwell equations. In this study, we follow such a mathematical system, and then analyze the EM force and their field-effect. In addition, the P&P for this calculation are dependent on time only, since, in confined region, the source and its induction field are in the same position. The schematics of the divergence-equations $\nabla \cdot \mathbf{B}=0$ and $\nabla \cdot \mathbf{D}=\rho$ are shown in Figs. 2(a) and (b), respectively. The investigation is performed in a confined three-dimensional (3D) region with charge-sources, of which the magnetic- and electric-charge density equals to zero and $\rho$, respectively. Based on the assumption II, the divergence-equations are identical as the classical ones since there does not exist any partial derivative to time. The equations are given as:

$$\begin{cases} \nabla \cdot \mathbf{B} = 0 \\ \nabla \cdot \mathbf{D} = \rho \end{cases}. \tag{7}$$

Obviously, they are same as the classical Maxwell equations if let the P&P be time-independent. Let $\rho$ harmonic that $\rho=\exp(-j\omega t)$. We can get the results after taking the derivative of Eq. (7) to time (detailed in Supplementary material S3):

$$\begin{cases} \dfrac{\partial \nabla \cdot \mathbf{E}}{\partial t} + (A + j\omega)\nabla \cdot \mathbf{E} = 0 \\ \dfrac{\partial \nabla \cdot \mathbf{H}}{\partial t} + (A + j\omega)\nabla \cdot \mathbf{H} = 0 \end{cases}. \tag{8}$$

Eq. (8) demonstrates that the time-term of $\nabla \cdot \mathbf{E}$ and $\nabla \cdot \mathbf{H}$ should be $\exp[-(A+j\omega)t]$. So that, the time-term of **E** and **H** should be $\exp[-(A+j\omega)t]$ as well, since **E** and **H** are position-independent in the confined region. This indicates that the magnitude of **E** and **H**, in time-varied P&P case, decrease with time following exp(-*At*). Note that, the **E** or **H** magnitude here signifies the intensity of any arbitrary EM fields in the random position, i.e., it demonstrates the level EM intensity at the time.

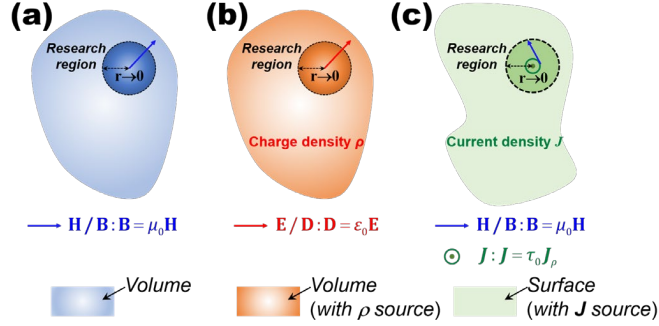

Fig. 2. Static electric and magnetic fields in confined region with source. (a) Schematic of $\nabla\cdot\mathbf{B}=0$. (b) Schematic of $\nabla\cdot\mathbf{B}=\rho$. The investigation is performed in a confined 3D space (the radius of the research region *r* tends to zero) with magnetic and electric charge density equaling to zero and $\rho$, respectively. (c) Schematic of $\nabla\times\mathbf{H}=\mathbf{J}$. The investigation is performed in a confined 3D curve (*r*→0) around current with density of **J**. The level intensity of the electric / magnetic fields and the effect current are decreased with time.

In the two curl-equations, the $\nabla\times\mathbf{E}$ equation is same as Eq. (1) since there exists no magnetic-current. However, the $\nabla\times\mathbf{H}$ equation should be corrected since the level intensity of **H** decreases with time. Fig. 2(c) shows the research schematic, that the investigation is performed in a confined region with a 3D curve around the current density **J**. Based on the classical Maxwell equations $\nabla\times\mathbf{H}=\mathbf{J}$ and the decrease term of **H** exp(-*At*), the $\nabla\times\mathbf{H}$ equation in time-varied P&P case is rewritten as (detailed in Supplementary material S3):

$$\begin{cases} \nabla\times\mathbf{H} = \dfrac{\partial}{\partial t}\mathbf{J} \\ \nabla\cdot\mathbf{J}_\rho = -\dfrac{\partial \rho}{\partial t} \\ \textbf{Constitutive relation: } \mathbf{J} = \tau_0 \mathbf{J}_\rho \\ \qquad\qquad\qquad \tau_0 = \tau_0^o e^{-At} \end{cases} \qquad (9)$$

Herein, **J** performs a new significance which presents the current-equivalence of the magnetic effect. A time-varied factor $\tau_0$ is introduced that $\mathbf{J}=\tau_0\mathbf{J}_\rho$, in which $\mathbf{J}_\rho$ is the current form based on the classical definition, namely the charge-motion. From Eq. (9), $\tau_0$ decreases with time, which indicates that, in time-varied P&P case, the magnetic effect of charge-motion is not invariable, but weakens with time. Note that, $\tau_0^o$ is introduced to present the general relationship between **J** and $\mathbf{J}_\rho$, which dose not has a defined value here. The value should be determined through measurements and a certain definition. In addition, in the time-varied P&P case, the level charge value is constant with time which is consistent with the principle of charge conservation. Its electric field-effect decrease is presented by the increase of $\varepsilon_0^o$. Whereas, the decrease of magnetic field-effect of current is presented by the decrease of effect current **J** and the increase of $\mu_0^o$, collectively.

Based on the classical mathematical system, the equations of static EM fields have been obtained. The physical results including the EM force and field-effect are herein discussed. In the confined region, a base system composed of two charges $q^*_1$ and $q^*_2$ with distance *r* is employed for the electric force analysis, as shown in Fig. 3(a), while a system composed of two infinitesimal currents $\boldsymbol{J}^*_1 dl$ and $\boldsymbol{J}^*_2 dl$ is employed for the magnetic force analysis, as shown in Fig. 3(b). The force between charges or currents can be given on the mathematical results above:

$$\begin{cases} \mathbf{F}^E_{12} = \dfrac{1}{4\pi\varepsilon^o_0 e^{At}} \dfrac{q^*_1 q^*_2}{r^2} \mathbf{e_r} \\ \mathbf{F}^M_{12} = \dfrac{\mu^o_0 e^{At}}{4\pi} \dfrac{\tau^o_0 e^{-At} \boldsymbol{J}^*_1 dl \times \left(\tau^o_0 e^{-At} \boldsymbol{J}^*_2 dl \times \mathbf{e_r}\right)}{r^2} \end{cases}. \quad (10)$$

where, $\mathbf{F}^E$ and $\mathbf{F}^M$ denote, respectively, the electric and magnetic force, with the subscripts 12 signifying the force of source2 from source1. $\mathbf{e_r}$ is the unit-vector pointing from source1 to source2. Eq. (10) indicates that both the electric and magnetic force decreases with time following $\exp(-At)$. This force also presents a level indicator at the time, since $q^*$ and $\boldsymbol{J}^*dl$ are arbitrarily selected. The weakening force implies that the level field-effect of charge and current weaken with time. These physical results are coincident with the predictions on the mathematical analysis above.

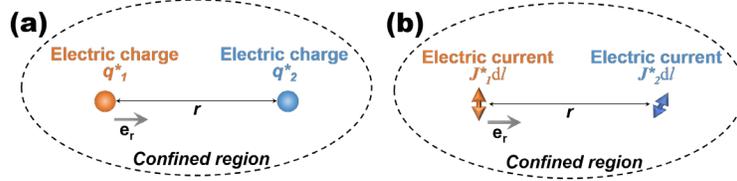

Fig. 3. Systems for static electric and magnetic force analysis. (a) Schematic of the electric force system. (b) Schematic of the magnetic force system. The base systems with dual-charge and dual-current are employed to discuss the electric and magnetic force, respectively. The EM force and field-effect are deceased with time.

Collectively, based on Eqs. (1), (7) and (9), the Maxwell equations of both static and covariant EM fields in the confined space-scale can be written as:

$$\begin{cases} \nabla \times \mathbf{H} = \mathbf{J} + \varepsilon^o_0 e^{At} \dfrac{\partial \mathbf{E}}{\partial t} + A\varepsilon^o_0 e^{At} \mathbf{E} \\ \nabla \times \mathbf{E} = -\left(\mu^o_0 e^{At} \dfrac{\partial \mathbf{H}}{\partial t} + A\mu^o_0 e^{At} \mathbf{H}\right) \\ \nabla \cdot \mathbf{B} = 0 \\ \nabla \cdot \mathbf{D} = \rho \\ \nabla \cdot \boldsymbol{J}_\rho = -\dfrac{\partial \rho}{\partial t} \\ \textbf{Constitutive relation:} \mathbf{D} = \varepsilon_0 \mathbf{E} \quad \mathbf{B} = \mu_0 \mathbf{H} \quad \mathbf{J} = \tau_0 \boldsymbol{J}_\rho \\ \qquad \qquad \qquad \quad \varepsilon_0 = \varepsilon^o_0 e^{At} \quad \mu_0 = \mu^o_0 e^{At} \quad \tau_0 = \tau^o_0 e^{-At} \end{cases}. \quad (11)$$

### 4.2 Electromagnetics in open space

In open space, there exist a spatial distance between the source and its induction field, which will result into a time-delay between the generation of and the EM field. This indicates that the P&P will shift during the field propagation. Therefore, the propagation velocity at each certain time should be achieved firstly in order to determine the relationship between the time-delay and spatial distance. In this study, the wave is assumed to propagate in the +z direction with $\mathbf{E}$ and $\mathbf{H}$ polarizing in the $\mathbf{x}$ and $\mathbf{y}$ directions, respectively. At any given time $t$, the $\varepsilon_0$ and $\mu_0$ can be considered as constants with values of $\varepsilon_0^o \exp(At)$ and $\mu_0^o \exp(At)$, respectively. In this case, the EM propagation is formulated as (detailed in Supplementary material S4):

$$\begin{cases} \mathbf{E} = \mathbf{e}_x E_0 e^{-j\omega\sqrt{\varepsilon_0^o \mu_0^o} e^{At} z - j\omega t} \\ \mathbf{H} = \mathbf{e}_y H_0 e^{-j\omega\sqrt{\varepsilon_0^o \mu_0^o} e^{At} z - j\omega t} \end{cases}, \qquad (12)$$

in which, $E_0$ and $H_0$ are the magnitude of electric and magnetic field at time $t$. The key parameters, including the characteristic-impedance $\eta(t)=|\mathbf{E}|/|\mathbf{H}|$, wavenumber $k(t)$, and propagation velocity $c(t)$, can be obtained:

$$\eta(t) = \sqrt{\frac{\mu_0^o}{\varepsilon_0^o}}, \quad k(t) = \omega\sqrt{\varepsilon_0^o \mu_0^o} e^{At}, \quad c(t) = \frac{1}{\sqrt{\varepsilon_0^o \mu_0^o} e^{At}}. \qquad (13)$$

Then, we first address the static EM field in the open space. Previously, the propagation features of static EM fields were not considered because the classical equations with sources, i.e., $\nabla \times \mathbf{H} = \mathbf{J}$ and $\nabla \cdot \mathbf{D} = \rho$, do not include any time factors in the time-independent P&P case. However, with time-varied P&P, the static EM characteristics should be discussed from the perspective of both time and position. The static electric and magnetic force between charges or currents are analyzed, achieving the result that the static EM force depends on the time only. Therefore, the P&P for describing the static EM fields are, correspondingly, dependent on the time only (detailed in Supplementary material S5). The static electric and magnetic field-effects of, respectively, charge and charge-motion decrease with time in the open space, which is identical as the results in the confined region. Because there exists no source in open space, the equations should be given in integral form. As shown in Fig. 4, $Sc$ is a surface bounded by a closed 3D curve $C$, and $Vs$ is a volume bounded by a closed surface $S$. $C$ and $S$ are selected arbitrarily. The equations are given as:

$$\begin{cases} \oint_C \mathbf{H} \cdot d\mathbf{l} = \iint_{Sc} \mathbf{J} \cdot d\mathbf{s} \\ \oint_C \mathbf{E} \cdot d\mathbf{l} = 0 \\ \oiint_S \mathbf{B} \cdot d\mathbf{s} = 0 \\ \oiint_S \mathbf{D} \cdot d\mathbf{s} = \oiiint_{Vs} \rho dv \\ \oiint_S \mathbf{J}_\rho \cdot d\mathbf{s} = -\frac{\partial}{\partial t} \oiiint_{Vs} \rho dv \\ \mathbf{Constitutive\ relation:} \mathbf{D} = \varepsilon_0 \mathbf{E} \quad \mathbf{B} = \mu_0 \mathbf{H} \quad \mathbf{J} = \tau_0 \mathbf{J}_\rho \\ \qquad\qquad \varepsilon_0 = \varepsilon_0^o e^{At} \quad \mu_0 = \mu_0^o e^{At} \quad \tau_0 = \tau_0^o e^{-At} \end{cases} \qquad (14)$$

The results above indicate that the curvilinear integral of $\mathbf{H}$ is independent on the curve selected or the current distribution, as shown in Fig. 4(a). The static magnetic phenomenon is completely determined by the P&P at the time. It is independent on position, even though the position difference may induce a time-delay which may further cause a P&P variation during the field propagation. As shown in Fig. 4(b), similar result can be achieved for the static electric field, of which the P&P depends on the time only. Collectively, for static EM field, the P&P in confined and open space are the same. Based on the Gauss and Stokes integral theorem, the static EM field can be described by Eq. (11) in the differential form, or, equivalently, by Eq. (14) in the integral form.

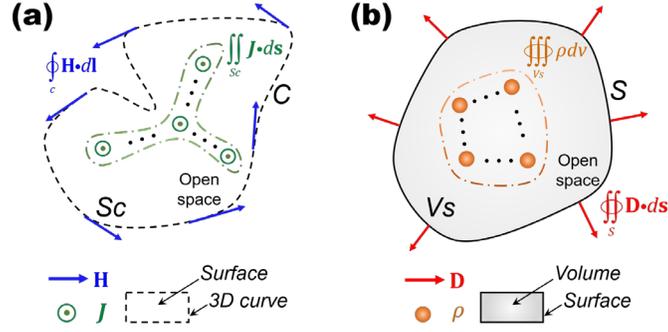

Fig. 4. Static electric and magnetic field in open space. (a) Schematic of equation $\oint \mathbf{H}\cdot d\mathbf{l}=\iint \mathbf{J}\cdot d\mathbf{s}$. (b) Schematic of equation $\oiint \mathbf{D}\cdot d\mathbf{s}=\iiint \rho dv$. The static EM force and field are independent on the curve / surface selected or the current / charge distribution. The P&P for regularizing static EM fields only depend on time.

We second address the covariant EM field in open space. Different form the static EM fields, the covariant EM field exists in the wave-form, namely the EM wave, which has two characteristics: (i) the electric and magnetic field are resonant, i.e., the alternating electric and magnetic fields induce each other at the same time and in the same position, so that the generation of and the induction field are regularized by the same P&P; (ii) the wave must be in propagation-state, i.e., the EM state shifts with both time and position, correspondingly, the P&P for regularizing the EM state will depend on position as well. In addition, such a position-dependence is caused by propagation, that one EM state can only impact the state in the next time, i.e., the closed next position, not the wave in any position. So that, the position-dependence of P&P is presented by their partial derivative to position, whereas, the value of permittivity, as well as permeability, is identical in all positions. As shown in Fig. 5(a), an EM wave propagates in +$\mathbf{z}$ direction. The wave at each place can be regarded as a single wave that propagates with the same characteristics, as shown in the right subgraph. For each single wave, the EM state at $t$ and ($t+\Delta t$) time are equivalent to that in, respectively, $z$ and ($z+\Delta z$) position, so that the P&P regularizing the EM state at $t$ and ($t+\Delta t$) time also regularize the state in, respectively, $z$ and ($z+\Delta z$) position.

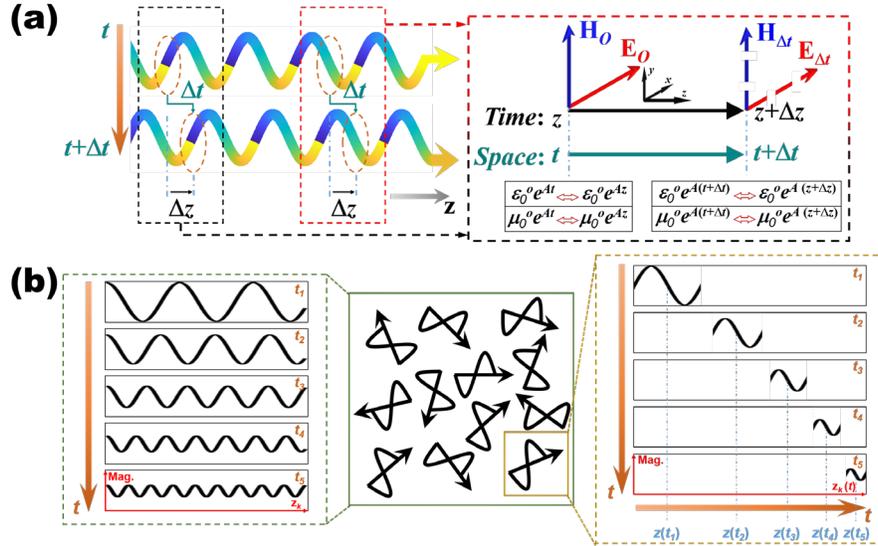

Fig. 5. Propagation electromagnetic wave with time-varied permittivity and permeability. (a) Schematic of the propagation of EM waves. The covariant EM filed propagates in wave form, so that the EM state shifts with the time and position simultaneously. Therefore, the P&P for regularizing the EM waves depends on both time and position. (b)

Two research models of EM wave. There exist EM waves with random wavevectors distributing random position. The propagation features of the general and single EM wave are shown in the left and right subgraph, respectively.

Based on the analysis above, a term $F(t, z)$ which depends on both time and position is introduced to define the P&P, meanwhile, its position-dependence should be presented by the partial derivative to position since $\Delta z$ should be tend to zero. The equations of covariant EM wave are given, with the partial derivatives of $F$ to $t$ and $z$ included, as (detailed in Supplementary material S4):

$$\begin{cases} \nabla \times \mathbf{H} = \varepsilon_0^o F \dfrac{\partial \mathbf{E}}{\partial t} + A\varepsilon_0^o F \mathbf{E} \\ \nabla \times \mathbf{E} = -\left( \mu_0^o F \dfrac{\partial \mathbf{H}}{\partial t} + A\mu_0^o F \mathbf{H} \right) \\ \nabla \cdot \mathbf{B} = 0 \\ \nabla \cdot \mathbf{D} = 0 \\ \textbf{Constitutive relation:} \mathbf{D} = \varepsilon_0 \mathbf{E} \quad \mathbf{B} = \mu_0 \mathbf{H} \\ \qquad \qquad \varepsilon_0 = \varepsilon_0^o F \quad \mu_0 = \mu_0^o F \\ \qquad \qquad F = e^{At} \quad \dfrac{\partial F}{\partial t} = AF \quad \dfrac{\partial F}{\partial z} = A\sqrt{\varepsilon_0^o \mu_0^o} F^2 \end{cases} \quad . \tag{15}$$

To solve the Eq. (15), the propagation can be described as:

$$\begin{cases} \mathbf{E} = \mathbf{e}_x E_0 e^{-j\left(\omega \sqrt{\varepsilon_0^o \mu_0^o} F\right) z - (A + j\omega) t} \\ \mathbf{H} = \mathbf{e}_y H_0 e^{-j\left(\omega \sqrt{\varepsilon_0^o \mu_0^o} F\right) z - (A + j\omega) t} \end{cases} \quad . \tag{16}$$

$F$ is introduced into the representations of P&P (in Eq. (15)) and the phase-term of EM wave (in Eq. (16)), achieving the coincident results between the constitutive relation and propagation features. Eq. (15) provides the general equations of covariant EM waves, in both the confined and open space. The Eq. (1) is only a special case of Eq. (15) that: Eq. (15) will transform into Eq. (1) when $\partial F/\partial z=0$, since the variation-with-position of P&P is unconsidered in the confined region.

Compare the equations of static EM fields and covariant waves, as presented in the Eqs. (11) and (15), respectively. They have the different constitutive relations, of which the derivatives of their P&P to position equal to zero and $A(\varepsilon_0^o \mu_0^o)^{1/2}F^2$, respectively. This is because the propagation of static EM field indicates an intensity variation only, which has been included in the exponential term (real term) of field representations. However, the propagation of the covariant wave indicates, additionally, a phase-varying, resulting into that an imaginary term (relevant to position, time and P&P) is introduced into the exponential term of the wave representations. So that, its P&P should regularize the phase variation with time and position simultaneously.

Two research models of EM wave are shown in Fig. 5(b), that the schematics of the general and single wave are shown in the left and right subgraph, respectively. There exist EM waves with random wavevectors distributing random position. The global features which follows the classical definition of EM art are described by Eq. (15), as shown in the left subgraph. The wavelength, wave velocity, and magnitude of all waves decrease with time. As shown in the right subgraph, since the P&P of free space are non-dispersive, an example of single-frequency wave is used to discuss the behavior of single EM wave from its generation to any time following. In this case, the binary function $F$ in Eq. (16) transforms into a unary function $F_s$, since the position $z$ is dependent on the time $t$. The propagation formulas are given as:

$$\begin{cases} \mathbf{E} = \mathbf{e_x} E_0^g e^{-j\left(\omega\sqrt{\varepsilon_0^g \mu_0^g} F_s\right)z - (A+j\omega)t} \\ \mathbf{H} = \mathbf{e_y} H_0^g e^{-j\left(\omega\sqrt{\varepsilon_0^g \mu_0^g} F_s\right)z - (A+j\omega)t} \\ F_s = e^{At} \\ z = \int_0^t \frac{1}{\sqrt{\varepsilon_0^g \mu_0^g} e^{At'}} dt' \end{cases}, \quad (17)$$

in which, $E_0^g$ and $H_0^g$ are the magnitude of, respectively, electric and magnetic field at the generation time. $\varepsilon_0^g$ and $\mu_0^g$ are the permittivity and permeability at the generation time, respectively. $F_s$ is the variation function of P&P, which is dependent on $t(z)$, namely $z(t)$. The relationship between $z$ and $t$ is shown in the last equation of Eq. (17). Comparison of Eqs. (16) and (17): (i) Eq. (16) is based on the general Cartesian coordinate which has the translational invariance, whereas, the coordinate of Eq. (17) exists a defined time-space relationship and a fixed origin; (ii) Eq. (16) describes the general waves, whereas, Eq. (17) describes one single wave at any time / position; (iii) the variation function $F$ is a binary function, of which the partial derivatives depend on time only, whereas, $F_s$ is a unary function, of which the derivative depends on time, as well as, position.

### 4.3 Electromagnetic energy and momentum

Based on the Maxwell equations of static and covariant EM fields, the EM energy representation, namely, the Poynting's theorem can be given as (detailed in Supplementary material S3):

$$\nabla \cdot (\mathbf{E} \times \mathbf{H}) = -\mathbf{E} \cdot \mathbf{J} - \frac{1}{2}\left(\frac{\partial \mathbf{D} \cdot \mathbf{E}}{\partial t} + \frac{\partial \mathbf{B} \cdot \mathbf{H}}{\partial t}\right) - \frac{1}{2}\left(\mathbf{E} \cdot \mathbf{E} \frac{\partial \varepsilon_0}{\partial t} + \mathbf{H} \cdot \mathbf{H} \frac{\partial \mu_0}{\partial t}\right). \quad (18)$$

The term on the left is the divergence of EM energy flux density, signifying the net energy flux of a spatial region. On the right, the first and second terms denote the gradient-to-time of the EM loss and stored energy, respectively. These three terms construct the representation of the classical Poynting's theorem. However, in the time-varied P&P case, a new energy form which is caused by the P&P variation is found, as shown in the third term on the right. In source-free case, $\nabla \cdot (\mathbf{E} \times \mathbf{H}) = 0$, (detailed in Supplementary material S3) which indicates that the energy entering and leaving a source-free region is equal. This because the free space cannot provide any divergence-source of EM energy, that, mathematically, the EM intensity at different positions varies with time following the same formula $\exp(-At)$ in time-varied P&P case. Furthermore, there exists no ohmic loss, so $\mathbf{E} \cdot \mathbf{J} = 0$. In this case, the sum of the second and third terms equals to zero, which is accordant with the mathematical result based on that: the magnitude of $\mathbf{E}$ and $\mathbf{H}$ decrease with time following $\exp(-At)$, whereas the P&P increase following $\exp(At)$. Meanwhile, $A>0$, therefore, Eq. (18) indicates that the stored EM energy transforms into a new energy form in time-varied P&P case. The transformations of electric and magnetic energy are synchronous. These results of EM energy accord with the law of conservation of energy and the entropy increase principle.

Then, we discuss the EM momentum. In time-varied P&P case, the EM momentum density is also written as $\mathbf{D} \times \mathbf{B}$, the EM momentum theorem is presented as (detailed in Supplementary material S6):

$$\nabla \cdot \bar{\bar{\Psi}}_C + \nabla \cdot \left(\bar{\bar{\Psi}}_E + \bar{\bar{\Psi}}_M\right) = -\frac{\partial(\mathbf{D} \times \mathbf{B})}{\partial t}$$

$$\begin{cases} \bar{\bar{\Psi}}_C = \frac{1}{2\varepsilon_0}D^2\bar{\bar{I}} + \frac{1}{2\mu_0}B^2\bar{\bar{I}} - \frac{1}{\varepsilon_0}\mathbf{DD} - \frac{1}{\mu_0}\mathbf{BB} \\ \nabla \cdot \bar{\bar{\Psi}}_E = (\mathbf{D} \cdot \mathbf{D})\nabla\left(\frac{1}{2\varepsilon_0}\right) \quad \nabla \cdot \bar{\bar{\Psi}}_M = (\mathbf{B} \cdot \mathbf{B})\nabla\left(\frac{1}{2\mu_0}\right) \end{cases} \quad (19)$$

in which, $\Psi_C$ is the classical momentum-flow-density tensor. $\Psi_E$ and $\Psi_M$ are two new momentum-flow-density forms, which are caused by the derivative of $F$ to position in the P&P expression, further in physics, caused by the propagation character of the covariant EM field. In free-space, the magnitude of $\mathbf{E}$ and $\mathbf{H}$ decreases following $\exp(-At)$, while the P&P increases following $\exp(At)$. Therefore, the right term $\partial(\mathbf{D}\times\mathbf{B})/\partial t=0$, which indicates that the EM momentum is also conservative in time-varied P&P case. The difference from the classical momentum form is that two momentum-flow-density forms $\Psi_E$ and $\Psi_M$, i.e., two EM force forms $1/2(\mathbf{D}\bullet\mathbf{D})\nabla\varepsilon_0$ and $1/2(\mathbf{B}\bullet\mathbf{B})\nabla\mu_0$ are found here. They function as the divergence sources of EM momentum in the time-varied P&P case.

### 4.4 Propagation characters of electromagnetic wave

The propagation characters of EM wave have been presented in Eqs. (16) and (17). Herein, the key propagation indicators are summarized to exhibit their variation-with-time feature. As shown in Fig. 6, the wave is generated at time $t^g$, and its propagation indicators vary with $t$. For clear introduction, all the indicators are normalized by the values at the generation time. Fig. 6(a) shows the characteristic parameters of vacuum, including the vacuum permittivity $\mu_0$, vacuum permeability $\varepsilon_0$ and characteristic impedance $\eta$. $\mu_0$ and $\varepsilon_0$ increase with time following $\exp(At)$, and $\eta$ which equals to $\mathbf{E}/\mathbf{H}$, as well as equals to $(\mu_0/\varepsilon_0)^{1/2}$, is constant. The invariability of $\eta$ indicates that there exists no EM reflection during the wave propagation. Fig. 6(b) shows the propagation features, including the wavenumber $k$, wave velocity (lightspeed) $c$ and wavelength $\lambda$. $\lambda$ and $c$ decrease with time following $\exp(-At)$, conversely, $k$ increases following $\exp(At)$. Fig. 6(c) shows the intensity features, including the magnitude of electric and magnetic fields ($\mathbf{E}$ and $\mathbf{H}$) denoted by $M_E$ and $M_H$ respectively, the electric and magnetic stored-energy denoted by $Ene_E$ and $Ene_M$ respectively, and the EM momentum $Mom$. $M_{E\,H}$ decrease following $\exp(-At)$, which indicates that the level EM intensity decrease with time. $Ene_E$ and $Ene_M$ decrease following $\exp(-At)$, which indicates that the EM energy is transformed into a new energy form caused by the time-varied P&P. Furthermore, the energy decrease following the same formula as magnitude is because that $Ene_E$ and $Ene_M$ equal to, respectively, $(\mathbf{D}\bullet\mathbf{E})/2$ and $(\mathbf{B}\bullet\mathbf{H})/2$, in which only $\mathbf{E}$ and $\mathbf{H}$ decrease with time. $Mom$ is kept with time, however, there exist a new divergence source of EM momentum, i.e., a new EM force form which is cause by the propagation feature of EM wave.

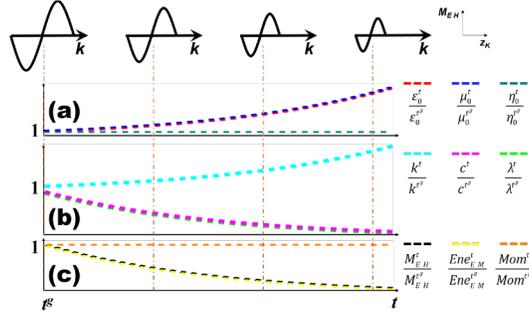

Fig. 6. Propagation characters of electromagnetic wave. (a) Variation of characteristic parameters of vacuum. (b) Variation of EM propagation features. (c) Variation of EM intensity features.

## 5. Comparisons between theorical results and observations

### 5.1 Electromagnetics based on observation

The analysis above introduces the $\varepsilon_0$, $\mu_0$ and $\tau_0$ to define the proportion relations of **D-E**, **B-H**, and **J-$J_\rho$** in the time-varied P&P case, respectively. They were the relation factors without certain values in the analysis above. Herein, the values of $\varepsilon_0$, $\mu_0$ and $\tau_0$ can be defined by the observation values based on the measurements at our time. Their observation values are denoted by $\varepsilon_0^{ob}$, $\mu_0^{ob}$ and $\tau_0^{ob}$, respectively. Obviously, the measurements should be based on some standardized definitions, such as on the International System of Units (SI). First, the force between effective-currents $\boldsymbol{J}^*$ is measured, to further define the value of $\boldsymbol{J}^*$ and $\mu_0^{ob}$. Second, let the defined $\boldsymbol{J}^*$ equal to the $\boldsymbol{J}_\rho^*$, since, in our general definition system, $\boldsymbol{J}^*$ is regarded as equivalent to $\boldsymbol{J}_\rho^*$. The charge-motion was actually defined by the magnetic effect $\boldsymbol{J}^*$. In this case, $\tau_0^{ob}$ is equivalent to one, since it has been included into the $\mu_0^{ob}$. Third, define the electric charge $q^*$ through the integration of $\boldsymbol{J}_\rho^*$. At last, the value of $\varepsilon_0^{ob}$ can be achieved by measuring the quantity of electric charges and the electric force between them (detailed in Supplementary material S7). Therefore, the equations of static EM fields can be given as:

$$\begin{cases} \nabla \times \mathbf{H} = \mathbf{J} \\ \nabla \times \mathbf{E} = 0 \\ \nabla \cdot \mathbf{B} = 0 \\ \nabla \cdot \mathbf{D} = \rho \\ \nabla \cdot \mathbf{J}_\rho = -\dfrac{\partial \rho}{\partial t} \\ \textbf{Constitutive relation:} \mathbf{D} = \varepsilon_0 \mathbf{E} \quad \mathbf{B} = \mu_0 \mathbf{H} \quad \mathbf{J} = \tau_0 \mathbf{J}_\rho \\ \quad \varepsilon_0 = \varepsilon_0^{ob} e^{At} \quad \mu_0 = \mu_0^{ob} e^{At} \quad \tau_0 = e^{-At} \end{cases} \quad , \quad (20)$$

and the equations of covariant EM fields can be given as:

$$\begin{cases} \nabla \times \mathbf{H} = \varepsilon_0^{ob} F \dfrac{\partial \mathbf{E}}{\partial t} + A\varepsilon_0^{ob} F \mathbf{E} \\ \nabla \times \mathbf{E} = -\left( \mu_0^{ob} F \dfrac{\partial \mathbf{H}}{\partial t} + A\mu_0^{ob} F \mathbf{H} \right) \\ \nabla \cdot \mathbf{B} = 0 \\ \nabla \cdot \mathbf{D} = 0 \\ \textbf{Constitutive relation:} \mathbf{D} = \varepsilon_0 \mathbf{E} \quad \mathbf{B} = \mu_0 \mathbf{H} \\ \quad \varepsilon_0 = \varepsilon_0^{ob} F \quad \mu_0 = \mu_0^{ob} F \\ \quad F = e^{At} \quad \dfrac{\partial F}{\partial t} = AF \quad \dfrac{\partial F}{\partial z} = A\sqrt{\varepsilon_0^{ob} \mu_0^{ob}} F^2 \end{cases} \quad . \quad (21)$$

In Eqs. (20) and (21), the value of $\varepsilon_0^{ob}$ and $\mu_0^{ob}$ equal to, respectively, $8.854187818 \times 10^{-12}$ C$^2$N$^{-1}$m$^{-2}$ and $4\pi \times 10^{-7}$ NA$^{-2}$. In the units, C (Coulomb), N (Newton), m (meter), and A (Ampere) are the unit of electric charge, force, length, and current, respectively.

### 5.2 Comparison between results on this theory and some observation results

The Eqs. (20) and (21) present the EM behavior after our time, since the $t$ should value positive. However, as shown in Fig. 7(a), the EM vision in the $t<0$ case can be observed by us, which happened in the past and with a distance from us. An EM wave was generated $T$ time ago, and then transmitted after an $R$-length journey to the observer, therefore, the observer can obtain the EM vision $R$ distance away. Here, a time parameter $t_n$ is introduced to denote the negative time, to present this EM journey. The introduction of $t_n$ can also differ from the positive time $t$, for the purpose that this discussion can present the results along both the positive and negative time-directions. Note that, $t_n$ is set bigger than zero for easy calculation, that the time in the EM journey is denoted by $-t_n$. In addition, since the discussion is based on the EM parameters at the observation time, the observation and generation time are set to $t_n=0$ and $t_n=-T$, respectively. Based on the analysis above, the wave velocity $c(t_n)$ and the distance $R(T)$ equals to:

$$\begin{cases} c(t_n) = \dfrac{1}{\sqrt{\varepsilon_0^{ob} \mu_0^{ob} e^{At} e^{-At_n}}} \\ R(T) = \displaystyle\int_0^T \dfrac{1}{\sqrt{\varepsilon_0^{ob} \mu_0^{ob} e^{At} e^{-At_n}}} dt_n \end{cases} \quad (22)$$

$T$ is actually a distance indicator, determining the distance between the generation and observation. $t$ goes forward indicating any observation time in the future, and $t_n$ goes backward indicating the transmission of the observed vision. Obviously, the lightspeed at the observation time $c^{ob}$ equals to $1/(\varepsilon_0^{ob} \mu_0^{ob})^{1/2}$.

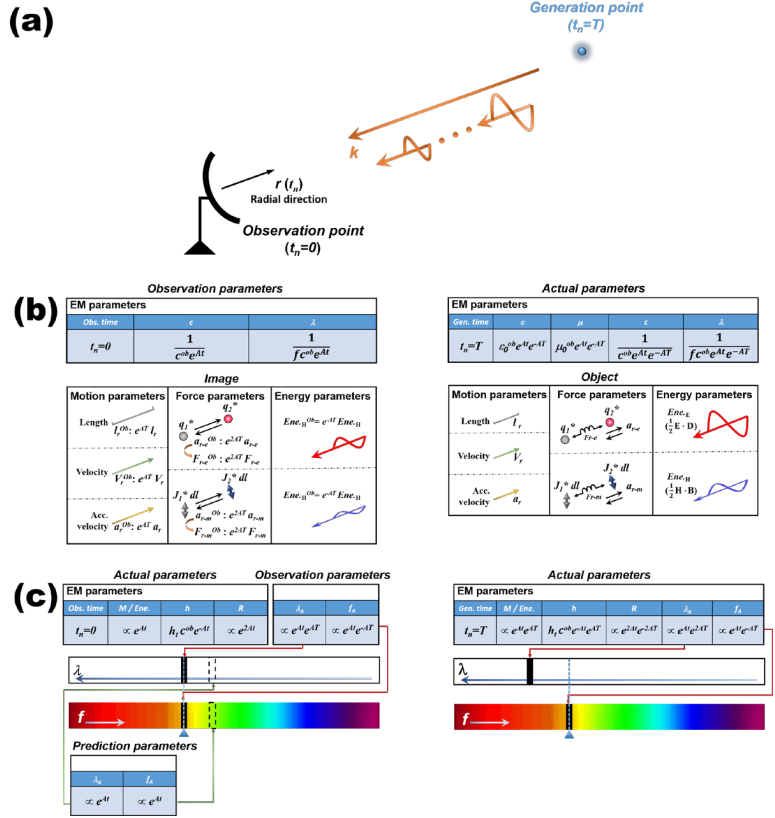

Fig. 7. Observed electromagnetic visions in the past. (a) Schematic of observation of EM vision in the past. The observer can obtain the EM vision $R$ distance away, which was generated at $T$ time ago. (b) Comparisons of the parameters at the generation ($t_n=T$) and observation time ($t_n=0$). The electromagnetism, motion, force and energy features in the radical direction are presented. (c) Spectrum variation of the observation. The absorption frequency in the observed spectrum exist a red-shift phenomenon.

Fig. 7(b) exhibits the parameters at the generation ($t_n=T$) and observation time ($t_n=0$), including the electromagnetism, motion, force and energy parameters. In this research, only the features in the radial direction are addressed, since only the radial light can be obtained by the observer. The wave is generated at the time with lower P&P, which has a higher lightspeed and a longer wavelength. The lightspeed and wavelength go down when the wave arrives the observation point, since they determined completely by the P&P at the observation time which increase by $\exp(AT)$. The observed EM view is an optical image of the objects at the generation time, i.e., the observed objects can be regarded as being immersed in the "water" with lower P&P. Therefore, all the imaged length, motion velocity and accelerated velocity are proportional to the lightspeed at the generation time, which are bigger than their actual values. Note that, such a length inflation is caused by the optical imaging which is essentially a refraction phenomenon. The EM force parameters can only be deduced by measuring the accelerated velocity between charges or currents, of which the accelerated velocity is determined by two factors: the force at the generation time and the observed length inflation. Both of them are proportional to $\exp(AT)$, so that, the observed force is proportional to $\exp(2AT)$. At last, the obtained electric and magnetic energy are both decreased, i.e., the actual EM energy is higher than the observed values.

The spectrum analysis is an effective method for cosmology research, which is addressed here to present the spectrum variation in the time-varied P&P case, as shown in Fig. 7(c). First, the Planck constant $h$ and Rydberg constant $R$ are investigated. As analyzed above, the level energy of EM fields and waves decreases with time, so there is reason to suppose that the energy per frequency (d$\nu$, $\nu$ denotes frequency as $f$, which is introduced here since it is a general representation in quantum mechanics) is bigger at the previous time. Based on the equation $Ene.=h\nu$, $h$ will decrease with time following $\exp(-At)$ as well in the time-varied P&P case. This demonstrates $h$ is dependent on the lightspeed $c$. We introduce a constant factor $h_t$ to let $h=h_t c$, that, based on $Ene.=hc/\lambda$, $Ene.=(h_t/\lambda)c^2$, in which $(h_t/\lambda)$ is found with the quality dimensionality. Furthermore, based on Ref. [3], the Rydberg constant $R$ is proportional to $\exp(At)\exp(-2AT)$, so that, in the spectrum, the absorption frequency and wavelength is proportional to $\exp(At)\exp(-AT)$ and $\exp(At)\exp(2AT)$ at the generation time, respectively. After the transmission, the frequency kept at $\exp(At)\exp(-AT)$ whereas the wavelength decreases to $\exp(At)\exp(AT)$. On the other hand, based on the Planck constant and Rydberg constant at the observation time, the prediction value of the absorption frequency and wavelength are both proportional to $\exp(At)$.

Collectively, first, there exist an inflation of length, velocity and accelerated velocity in the observed image. Based on the inflation coefficient $\exp(AT)$, an effective velocity leaving from the observer can be calculated, if we assume the lightspeed at the generation time equals to that at the observation time. The effective velocity $v_{eff\_im}$ can be achieved on the relationship $(v_{eff\_im}+c^{ob})/c^{ob}=\exp(AT)$, that:

$$v_{eff\_im} = c^{ob}e^{AT} - c^{ob}. \tag{23}$$

Second, the absorption spectrum is red-shift, of which the radio of the observed absorption frequency to the prediction frequency based on $h$ and $R$ at the observation time equal to $\exp(AT)$. If we if treat this red-shift as a Doppler effect, the effective leaving velocity $v_{eff\_sp}$ can be obtained on the relationship $1/(1+v_{eff\_sp}/c^{ob})= \exp(AT)$, that:

$$v_{eff\_sp} = c^{ob}e^{AT} - c^{ob}. \tag{24}$$

From Eqs. (23) and (24), the effective leaving velocity achieved on the two method are the same. The well-known Hubble's law demonstrates a proportional relationship between the effective leaving velocity and distance. Herein, we calculate, based on results with time-varied P&P, the radio of the leaving velocity $v_{eff\_im\_sp}$ and the distance $R$ is given as:

$$\frac{c^{ob}e^{AT} - c^{ob}}{\int_0^T c^{ob}e^{At'}dt'} = A. \tag{25}$$

Eq. (25) indicates that there exists a proportional relationship between $v_{eff\_im\ sp}$ and $R$ as well. Compare with the Hubble's law, the variation factor of P&P $A$ may equal to the Hubble constant, which have the same dimensionality of 1/s.

## 6. Conclusion

To summarize, based on the assumption that the vacuum P&P are time-varied, the Maxwell equation of static and covariant EM field are presented, in which the P&P increase with time exponentially. The EM features in time-varied P&P case are discussed, that (i) the vacuum lightspeed, level EM intensity, and magnetic effect of charge-motion decrease with time; (ii) the EM energy transforms into a new energy form caused by the time-varied P&P; (iii) a new momentum-tensor form, i.e., a new EM force-form is found in the covariant EM field. The time-dependent variation of P&P also results into the decrease-with-time of Planck constant and the increase of Rydberg constant. A red-shift of the absorption spectrum can be predicted, which is coincident with the Hubble's observation results.

# Supplementary materials: Electromagnetics with time-varied vacuum permittivity and permeability

ZHIWEI SUN,[1]

[1] School of Microelectronics and Communication Engineering, Chongqing University, Chongqing, 400044, China

**Abstract:** Maxwell equations provide a complete description of the electromagnetic (EM) phenomena, which have been one of the key fundamental-theories of modern physics, such as electromagnetism, optics, quantum theories, etc. The vacuum permittivity and permeability (P&P) in the constitutive relation were regarded as constant, resulting into that the vacuum lightspeed is constant with time. However, neither the Maxwell equations nor any experiments demonstrated that the P&P must be invariable with time. Here, the P&P are assumed as time-varied, achieving a mathematical result exhibits that the P&P are increased with time exponentially. Then, the Maxwell equations of static and covariant EM fields are presented, which include the different constitutive relations. The EM characters in the time-varied case are investigated that, the lightspeed, level EM intensity, and magnetic effect of charge-motion decreases with time. The EM energy transforms into a new energy form caused by the P&P variation, and a new momentum-tensor form caused by propagation is found as well. At last, the Maxwell equations based on the observation results at our time are presented. Meanwhile, since the level EM intensity decreases with time, the energy per frequency is time-decreased, resulting into that the Planck constant decreases exponentially, and correspondingly, the Rydberg constant increases with time. A red-shift of absorption spectrum is predicted based on this finding, which is coincident with the Hubble's observation results.

**S1: Variation functions of permittivity and permeability**

The functions of permittivity and permeability (P&P) can be determined based on the assumptions I to III. Herein, we detail the derivations of the definition of P&P function. Two relevant Eqs. are listed for a better introduction. First, the Maxwell equations of covariant EM field in the confined region are given as:

$$\begin{cases} \nabla \times \mathbf{H} = \dfrac{\partial \mathbf{D}}{\partial t} = \varepsilon_0 \dfrac{\partial \mathbf{E}}{\partial t} + \mathbf{E}\dfrac{\partial \varepsilon_0}{\partial t} \\ \nabla \times \mathbf{E} = -\dfrac{\partial \mathbf{B}}{\partial t} = -\mu_0 \dfrac{\partial \mathbf{H}}{\partial t} - \mathbf{H}\dfrac{\partial \mu_0}{\partial t} \\ \nabla \cdot \mathbf{B} = 0 \\ \nabla \cdot \mathbf{D} = 0 \\ \textbf{Constitutive relation: } \mathbf{D} = \varepsilon_0 \mathbf{E} \quad \mathbf{B} = \mu_0 \mathbf{H} \\ \qquad\qquad\qquad\qquad \varepsilon_0 = \varepsilon_0(t) \quad \mu_0 = \mu_0(t) \end{cases} \quad . \tag{S1}$$

and second, the Helmholtz equations can be obtained by the $\nabla \times (\nabla \times \mathbf{H})$ and $\nabla \times (\nabla \times \mathbf{E})$ calculations, which are given as:

$$\begin{cases} -\dfrac{\partial \varepsilon_0}{\partial t}\left(\dfrac{\partial \mu_0}{\partial t}\mathbf{H} + \mu_0 \dfrac{\partial \mathbf{H}}{\partial t}\right) - \varepsilon_0 \left(\dfrac{\partial^2 \mu_0}{\partial t^2}\mathbf{H} + 2\dfrac{\partial \mu_0}{\partial t}\dfrac{\partial \mathbf{H}}{\partial t} + \mu_0 \dfrac{\partial^2 \mathbf{H}}{\partial t^2}\right) + \nabla^2 \mathbf{H} = 0 \\ -\dfrac{\partial \mu_0}{\partial t}\left(\dfrac{\partial \varepsilon_0}{\partial t}\mathbf{E} + \varepsilon_0 \dfrac{\partial \mathbf{E}}{\partial t}\right) - \mu_0 \left(\dfrac{\partial^2 \varepsilon_0}{\partial t^2}\mathbf{E} + 2\dfrac{\partial \varepsilon_0}{\partial t}\dfrac{\partial \mathbf{E}}{\partial t} + \varepsilon_0 \dfrac{\partial^2 \mathbf{E}}{\partial t^2}\right) + \nabla^2 \mathbf{E} = 0 \end{cases}. \quad (S2)$$

The denotations in Eqs. (S1) and (S2) can refer to the definitions in Eqs. (1) and (2) in the manuscript. Based on the assumption III, the coefficient of $\mathbf{H}$, $\partial \mathbf{H}/\partial t$ and $\partial \mathbf{H}^2/\partial^2 t$ in Eq. (S2) should be equal to that of $\mathbf{E}$, $\partial \mathbf{E}/\partial t$ and $\partial \mathbf{E}^2/\partial^2 t$, respectively. So that:

$$\begin{cases} \dfrac{\partial \varepsilon_0}{\partial t}\dfrac{\partial \mu_0}{\partial t} + \varepsilon_0 \dfrac{\partial^2 \mu_0}{\partial t^2} = \dfrac{\partial \mu_0}{\partial t}\dfrac{\partial \varepsilon_0}{\partial t} + \mu_0 \dfrac{\partial^2 \varepsilon_0}{\partial t^2} \\ \mu_0 \dfrac{\partial \varepsilon_0}{\partial t} + 2\varepsilon_0 \dfrac{\partial \mu_0}{\partial t} = \varepsilon_0 \dfrac{\partial \mu_0}{\partial t} + 2\mu_0 \dfrac{\partial \varepsilon_0}{\partial t} \\ \varepsilon_0 \mu_0 = \mu_0 \varepsilon_0 \end{cases}. \quad (S3)$$

To solve Eq. (S3), we can get:

$$\begin{cases} \varepsilon_0 = \varepsilon_0^o e^{f(t)} \\ \mu_0 = \mu_0^o e^{f(t)} \end{cases}, \quad (S4)$$

in which, $f(t)$ is an arbitrarily function of time. First, based on the assumption III and Eq. (S1), $f(t)$ should be a real function. Second, $df(t)^2/d^2t$ should equal to zero for three reasons. (i) Plug Eq. (S4) into Eq. (S2), we can, taking the first Eq. as example, get:

$$-\varepsilon_0^o \mu_0^o \dfrac{df}{dt} e^{2f} \left[\dfrac{df}{dt}\mathbf{H} + \dfrac{\partial \mathbf{H}}{\partial t}\right] - \varepsilon_0^o \mu_0^o e^{2f} \left\{\left[\left(\dfrac{df}{dt}\right)^2 + \dfrac{d^2 f}{dt^2}\right]\mathbf{H} + 2\dfrac{df}{dt}\dfrac{\partial \mathbf{H}}{\partial t} + \dfrac{\partial^2 \mathbf{H}}{\partial t^2}\right\} + \nabla^2 \mathbf{H} = 0. \quad (S5)$$

Eq. (S5) exists the $df(t)^2/d^2t$ term, whereas, the Maxwell equations in Eq. (S1) only has the $f(t)$ and $df(t)/dt$ terms. So that, the Maxwell equations cannot cover the information of electromagnetic (EM) phenomena if $df(t)^2/d^2t \neq 0$. (ii) Plug the solution of Eq. (S5) into Eq. (S1), the vacuum characteristic impedance which equals to $|\mathbf{E}|/|\mathbf{H}|$ should be a function-of-time rather than a constant if $df(t)^2/d^2t \neq 0$. This will, first, violate the assumption III, and second will induce an EM reflection during the EM propagation. Obviously, there must exist no EM reflection in vacuum, since we never observed an EM wave from past us. (iii) Based on the analysis of Eqs. (23) to (25) in the manuscript, $v_{eff}/R$ will not be a constant if $df(t)^2/d^2t \neq 0$. Collectively, $df(t)^2/d^2t$ should be equal to zero, therefore, Eq. (S4) is written as:

$$\begin{cases} \varepsilon_0 = \varepsilon_0^o e^{At} \\ \mu_0 = \mu_0^o e^{At} \end{cases}. \quad (S6)$$

**S2: Energy-flux analysis with time-varied permittivity and permeability**

The EM energy-flux analysis, i.e., the Poynting's theorem in time-varied P&P case is also based on the energy relationship between mechanical energy density $w_p$ and EM energy density $w_{em}$:

$$\dfrac{dw_p}{dt} = -\dfrac{dw_{em}}{dt} = \mathbf{J} \cdot \mathbf{E}, \quad (S7)$$

in which, $J$ is the effect current-density. This Eq. demonstrates an energy conversion between the mechanical energy and EM energy. First, for the static EM fields, the relevant Eqs. are list:

$$\begin{cases} \nabla \times \mathbf{H} = \mathbf{J} \\ \nabla \times \mathbf{E} = 0 \\ \nabla \cdot \mathbf{J}_\rho = -\dfrac{\partial \rho}{\partial t} \\ \textbf{Constitutive relation:} \mathbf{D} = \varepsilon_0 \mathbf{E} \quad \mathbf{B} = \mu_0 \mathbf{H} \quad \mathbf{J} = \tau_0 \mathbf{J}_\rho \\ \qquad \qquad \qquad \qquad \varepsilon_0 = \varepsilon_0^o e^{At} \quad \mu_0 = \mu_0^o e^{At} \quad \tau_0 = \tau_0^o e^{-At} \end{cases} \tag{S8}$$

Add $\mathbf{E}\cdot\nabla\times\mathbf{H}$ and $-\mathbf{H}\cdot\nabla\times\mathbf{E}$, we can get:

$$\nabla \cdot (\mathbf{E} \times \mathbf{H}) = [\mathbf{E} \cdot \nabla \times \mathbf{H} - \mathbf{H} \cdot \nabla \times \mathbf{E}] = -\mathbf{E} \cdot \mathbf{J}, \tag{S9}$$

in which, the left term $\nabla\cdot(\mathbf{E}\times\mathbf{H})$ denotes the divergence source of energy flux density in the space-dimension, and the right term $-\mathbf{E}\cdot\mathbf{J}$ denotes the energy change-gradient in the time-dimension. The Poynting vector $\mathbf{S}$ in time-varied P&P case is also given as $\mathbf{S}=\mathbf{E}\times\mathbf{H}$. Fig. S1 (a) shows the sourced case in which $\mathbf{J}\neq 0$. From Eq. (S9), $\mathbf{J}$ should be decease as well following exp(-$At$), since both $\mathbf{E}$ and $\mathbf{H}$ deceases with time. This, from the aspect of energy, validates that the effect-current is deceased with time. In the source-free case as shown in Fig. S1 (b), the free-space cannot provide the divergence source, i.e., the income and outgo EM energy flux are equivalent at any certain time. Collectively, for the static EM field, the physical (decrease of $\mathbf{E}$, $\mathbf{H}$ and $\mathbf{J}$) and mathematical (Eq. (S9)) results are coincident.

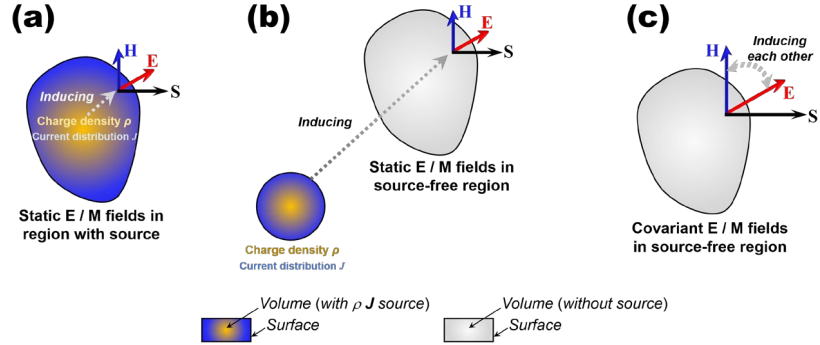

Fig. S1. Schematic of electromagnetism energy-flux analysis. (a) Static EM fields in sourced region. (b) Static EM fields in source-free region. $\mathbf{E}$, $\mathbf{H}$ and $\mathbf{J}$ are all decrease with time, which coincides with the mathematical results of EM energy analysis. (c) Covariant EM fields in source-free case. The EM stored-energy transforms into a new energy form which is caused by the time-varied P&P.

The covariant EM fields in source-free case is shown in Fig. S1 (c), in which the electric and magnetic field induce each other mutually. The relevant Eqs. are:

$$\begin{cases} \nabla \times \mathbf{H} = \dfrac{\partial \mathbf{D}}{\partial t} = \varepsilon_0 F \dfrac{\partial \mathbf{E}}{\partial t} + A\varepsilon_0 \mathbf{E} \\ \nabla \times \mathbf{E} = -\dfrac{\partial \mathbf{B}}{\partial t} = -\left(\mu_0 \dfrac{\partial \mathbf{H}}{\partial t} + A\mu_0 \mathbf{H}\right) \\ \textbf{Constitutive relation:} \mathbf{D} = \varepsilon_0(t,z)\mathbf{E} \quad \mathbf{B} = \mu_0(t,z)\mathbf{H} \\ \quad\quad \varepsilon_0 = \varepsilon_0^o F \quad \mu_0 = \mu_0^o F \\ \quad\quad F = e^{At} \quad \dfrac{\partial F}{\partial t} = AF \quad \dfrac{\partial F}{\partial z} = A\sqrt{\varepsilon_0^o \mu_0^o} F^2 \end{cases} \quad . \quad (S10)$$

At first, detail the Eq. (5) in the manuscript. In the confined region, the propagation feature is not considered, while the P&P are independent on position $\partial F/\partial z=0$. In this case, $\mathbf{E}$, $\mathbf{H}$, $\varepsilon_0$ and $\mu_0$ are the function of time only, i.e., only the derivative of $\mathbf{E}$, $\mathbf{H}$, $\varepsilon_0$ and $\mu_0$ to time needs to be considered. So that, $\nabla\bullet(\mathbf{E}\times\mathbf{H})=0$, of which $\nabla\bullet(\mathbf{E}\times\mathbf{H})$ can be written by adding $\mathbf{E}\bullet\nabla\times\mathbf{H}$ and $-\mathbf{H}\bullet\nabla\times\mathbf{E}$:

$$\nabla\bullet(\mathbf{E}\times\mathbf{H}) = -\frac{1}{2}\left(\varepsilon_0 \frac{\partial}{\partial t}E^2 + \mu_0 \frac{\partial}{\partial t}H^2\right) - \left(A\varepsilon_0 E^2 + A\mu_0 H^2\right). \quad (S11)$$

The first term on the right should be positive based on the assumption IV, that the EM energy can only transmit from EM fields to space. Therefore, the second right term must be negative. Because $E^2$ and $H^2$ are positive, $-A$ must be less than 0, i.e., $A>0$. The introduction of Eq. (S11) (as well as Eq. (5) in the manuscript) is because it can determine the sign of $A$ directly.

Second, solve the Poynting's theorem of covariant EM fields in the general case. We provide the derivation of $\mathbf{E}\bullet\partial\mathbf{D}/\partial t$ and $\mathbf{H}\bullet\partial\mathbf{B}/\partial t$ here, which are written as:

$$\begin{cases} \mathbf{E}\bullet\dfrac{\partial \mathbf{D}}{\partial t} = \dfrac{1}{2}\left(\dfrac{\partial \mathbf{D}\bullet\mathbf{E}}{\partial t} + \mathbf{E}\bullet\mathbf{E}\dfrac{\partial \varepsilon_0}{\partial t}\right) \\ \mathbf{H}\bullet\dfrac{\partial \mathbf{B}}{\partial t} = \dfrac{1}{2}\left(\dfrac{\partial \mathbf{B}\bullet\mathbf{H}}{\partial t} + \mathbf{H}\bullet\mathbf{H}\dfrac{\partial \mu_0}{\partial t}\right) \end{cases}, \quad (S12)$$

because:

$$\begin{cases} \mathbf{E}\bullet\dfrac{\partial \mathbf{D}}{\partial t} = \mathbf{E}\bullet\left(\varepsilon_0 \dfrac{\partial \mathbf{E}}{\partial t} + \mathbf{E}\dfrac{\partial \varepsilon_0}{\partial t}\right) = \mathbf{D}\bullet\dfrac{\partial \mathbf{E}}{\partial t} + \mathbf{E}\bullet\mathbf{E}\dfrac{\partial \varepsilon_0}{\partial t} = \dfrac{\partial \mathbf{D}\bullet\mathbf{E}}{\partial t} - \mathbf{E}\bullet\dfrac{\partial \mathbf{D}}{\partial t} + \mathbf{E}\bullet\mathbf{E}\dfrac{\partial \varepsilon_0}{\partial t} \\ \mathbf{H}\bullet\dfrac{\partial \mathbf{B}}{\partial t} = \mathbf{H}\bullet\left(\mu_0 \dfrac{\partial \mathbf{E}}{\partial t} + \mathbf{E}\dfrac{\partial \mu_0}{\partial t}\right) = \mathbf{B}\bullet\dfrac{\partial \mathbf{H}}{\partial t} + \mathbf{H}\bullet\mathbf{H}\dfrac{\partial \mu_0}{\partial t} = \dfrac{\partial \mathbf{B}\bullet\mathbf{H}}{\partial t} - \mathbf{H}\bullet\dfrac{\partial \mathbf{B}}{\partial t} + \mathbf{H}\bullet\mathbf{H}\dfrac{\partial \mu_0}{\partial t} \end{cases}. \quad (S13)$$

Add $\mathbf{E}\bullet(\nabla\times\mathbf{H}-\partial\mathbf{D}/\partial t)$ and $-\mathbf{H}\bullet(\nabla\times\mathbf{E}+\partial\mathbf{B}/\partial t)$, the Poynting's theorem of covariant EM fields can be given as:

$$\nabla\bullet(\mathbf{E}\times\mathbf{H}) = -\frac{1}{2}\left(\frac{\partial \mathbf{D}\bullet\mathbf{E}}{\partial t} + \frac{\partial \mathbf{B}\bullet\mathbf{H}}{\partial t}\right) - \frac{1}{2}\left(\mathbf{E}\bullet\mathbf{E}\frac{\partial \varepsilon_0}{\partial t} + \mathbf{H}\bullet\mathbf{H}\frac{\partial \mu_0}{\partial t}\right). \quad (S14)$$

At last, summarize the results of static and covariant EM fields. The general Poynting's theorem in the time-varied P&P can be obtained by summing Eqs. (S9) and (S14):

$$\nabla\bullet(\mathbf{E}\times\mathbf{H}) = -\mathbf{E}\bullet\mathbf{J} - \frac{1}{2}\left(\frac{\partial \mathbf{D}\bullet\mathbf{E}}{\partial t} + \frac{\partial \mathbf{B}\bullet\mathbf{H}}{\partial t}\right) - \frac{1}{2}\left(\mathbf{E}\bullet\mathbf{E}\frac{\partial \varepsilon_0}{\partial t} + \mathbf{H}\bullet\mathbf{H}\frac{\partial \mu_0}{\partial t}\right). \quad (S15)$$

In the open space without source, $\mathbf{E}\bullet\mathbf{J}$ equals to zero obviously. Meanwhile, $\nabla\bullet(\mathbf{E}\times\mathbf{H})$ in the general

case can be obtained based on Eqs. (S10), (S15), and the Eq. (16) in the manuscript:

$$\nabla \bullet (\mathbf{E} \times \mathbf{H}) = -j\omega \left( F - AF^2 \sqrt{\varepsilon_0^o \mu_0^o} \, z \right) \left[ \mu_0^o \mathbf{H} \bullet \mathbf{H} + \varepsilon_0^o \mathbf{E} \bullet \mathbf{E} \right]. \tag{S16}$$

Based on the classical energy research, the imaginary energy-variation has no significance, therefore, $\nabla \bullet (\mathbf{E} \times \mathbf{H})$ should be taken a real, which equals to zero.

**S3: Level magnitude of electric / magnetic fields and effect current**

Based on the assumption I, the EM features are independent on position, so that, the level magnitude of EM fields and the effect current can be investigated in any confined region. Obviously, the region should include charge and current sources. The relevant Eqs. are:

$$\begin{cases} \nabla \times \mathbf{H} = \mathbf{J} + \varepsilon_0^o e^{At} \dfrac{\partial \mathbf{E}}{\partial t} + A\varepsilon_0^o e^{At} \mathbf{E} \\ \nabla \times \mathbf{E} = -\left( \mu_0^o e^{At} \dfrac{\partial \mathbf{H}}{\partial t} + A\mu_0^o e^{At} \mathbf{H} \right) \\ \nabla \bullet \mathbf{B} = 0 \\ \nabla \bullet \mathbf{D} = \rho \\ \text{Constitutive relation:} \mathbf{D} = \varepsilon_0 \mathbf{E} \quad \mathbf{B} = \mu_0 \mathbf{H} \\ \qquad\qquad\qquad\qquad \varepsilon_0 = \varepsilon_0^o e^{At} \quad \mu_0 = \mu_0^o e^{At} \end{cases} \tag{S17}$$

Use $\nabla \bullet \mathbf{D} = \rho$ and let $\rho$ simple-harmonic that $\rho = \exp(-j\omega t)$. Take the derivative of $\nabla \bullet \mathbf{D} = \rho$ to $t$, we can get:

$$\frac{\partial \nabla \bullet \mathbf{E}}{\partial t} + (A + j\omega) \nabla \bullet \mathbf{E} = 0. \tag{S18}$$

Therefore, $\nabla \bullet \mathbf{E} = (\nabla \bullet \mathbf{E})_p \exp[-(A+j\omega)t]$, in which $(\nabla \bullet \mathbf{E})_p$ is a time-independent function. In the confined region, $\mathbf{E}$ is position-independent, therefore, $\mathbf{E}$ has same expression-of-time as $\nabla \bullet \mathbf{E}$. So that $\mathbf{E} = \mathbf{E}_p \exp[-(A+j\omega)t]$, in which $\mathbf{E}_p$ is also a time-independent function. This result can also be achieved by the equations of covariant EM fields, as provided in the Eq. (15) in the manuscript. The physical reason is that, for covariant EM fields, the level magnitude and energy are also determined by the P&P at that time only. The position-dependence of $F$ only affect the phase character, since such a dependence is caused by the propagation feature (i.e., the phase shift feature) of EM waves. Similarly, take the derivative of $\nabla \bullet \mathbf{B} = 0$ to $t$, we can get:

$$\frac{\partial \nabla \bullet \mathbf{H}}{\partial t} + (A + j\omega) \nabla \bullet \mathbf{H} = 0. \tag{S19}$$

So that, we obtain the result $\mathbf{H} = \mathbf{H}_p \exp[-(A+j\omega)t]$, in which $\mathbf{H}_p$ is a time-independent function. The Eq. $\nabla \times \mathbf{H} = \mathbf{J}$ is employed to discuss the effect current. Plug $\mathbf{H} = \mathbf{H}_p \exp[-(A+j\omega)t]$ into equation $\partial(\nabla \times \mathbf{H})/\partial t = \partial \mathbf{J}/\partial t$, we can get:

$$-(A + j\omega) \nabla \times \mathbf{H} = -(A + j\omega) \mathbf{J} = \frac{\partial}{\partial t} \mathbf{J}. \tag{S20}$$

So that, $\mathbf{J}$ should be written as $\mathbf{J}_p \exp[-(A+j\omega)t]$, where $\mathbf{J}_p$ is a time-independent function as well. Based on the assumption II, Eq.(S20) should be same as the result based on the classical Maxwell equations when let $A\rightarrow 0$. This means $\nabla\times\mathbf{H}=\mathbf{J}=\mathbf{J}_p$ should be identical to $\nabla\times\mathbf{H}=\mathbf{J}_\rho$, where $\mathbf{J}_\rho$ is classically defined by $\nabla\cdot\mathbf{J}_\rho=\partial\rho/\partial t$. Therefore $\mathbf{J}_p=\mathbf{J}_\rho$. Herein, for generally exhibiting the magnetic effect of current, a scaling factor $\tau_0$ is introduced additionally, that $\mathbf{J}$ is written as $\mathbf{J}=\tau_0\exp(-At)\mathbf{J}_\rho$. This indicates that the magnetic effect of charge-motion is decreased with time.

## S4: Propagation of covariant electromagnetic wave

The propagation velocity of EM wave in vacuum, namely the lightspeed, should be solved first. At a certain time $t_0$, the permittivity and permeability equal to $\varepsilon_0^o\exp(At_0)$ and $\mu_0^o\exp(At_0)$, respectively. They are both determined numbers. Hence, the propagation equations can be obtained based on the solving methods of the classical propagation equations, namely the solution of Helmholtz equations, which can be written as:

$$\begin{cases}\mathbf{E}=\mathbf{e}_x E_0(t_0)e^{-j\omega\sqrt{\varepsilon_0^o\mu_0^o}e^{At_0}z-j\omega t}\\ \mathbf{H}=\mathbf{e}_y H_0(t_0)e^{-j\omega\sqrt{\varepsilon_0^o\mu_0^o}e^{At_0}z-j\omega t}\end{cases}. \quad (S21)$$

The EM wave is discussed in the Cartesian coordinate, which radiates along $+z$ axis with $\mathbf{E}$ and $\mathbf{H}$ polarized in $\mathbf{x}$ and $\mathbf{y}$ directions, respectively. $E_0(t_0)$ and $H_0(t_0)$ are the level magnitude of, respectively, the $\mathbf{E}$ and $\mathbf{H}$ fields at time $t_0$. Eq. (20) indicates that the lightspeed at any general time $t$ equals to:

$$c(t)=\frac{1}{\sqrt{\varepsilon_0^o\mu_0^o}}e^{-At}. \quad (S22)$$

The EM wave has two characters, that: (i) the alternating electric and magnetic fields induce each other at the same time and in the same position, and (ii) the wave is in propagation state that the phase varied with both time ($\omega t$) and space ($kz$), therefore, each EM state is described by the P&P at one time as well as in the corresponding position. Collectively, (i) a position-dependence of P&P is induced by the propagation feature, and (ii) the EM state can only affect the state in the next time, i.e., the closed next position, correspondingly. So that, the variation-with-position of P&P for regularizing the state variation is performed by the partial derivative to position. The actual values of P&P are identical in all positions.

As shown in Fig. S2 (a), at one time, the propagation characters are uniform in every position, of which the EM characters at each position are shown in the right subgraph. In arbitrary position, the wave state in position $z$ at time $t$ will propagate into the next position ($z+\Delta z$) at the next time ($t+\Delta t$). This indicates that the P&P parameters for describing the propagation state should shift with both position and time. Herein, permittivity and permeability are assumed equating to $\varepsilon_0^o F$ and $\mu_0^o F$, respectively, where $F$ is a binary function of both position and time. Based on the assumption I, the permittivity and permeability equals to, respectively, $\varepsilon_0^o\exp(At)$ and $\mu_0^o\exp(At)$ at time $t$, i.e., $F=\exp(At)$. Based on the summary (ii) above (the EM state can only affect the state in the next time, i.e., the closed next position), $\Delta t$ and $\Delta z$ should tend to zero, so that the derivatives of $\varepsilon_0$ and $\mu_0$ to time and position can be solved based on the derivative-definition:

$$\begin{cases}\dfrac{\partial F}{\partial t}=\lim_{\Delta t\to 0}\dfrac{e^{A(t+\Delta t)}-e^{At}}{\Delta t}=e^{At}\lim_{\Delta t\to 0}\dfrac{e^{A\Delta t}-1}{\Delta t}=e^{At}\lim_{\Delta t\to 0}\dfrac{Ae^{A\Delta t}}{1}=Ae^{At}\\ \dfrac{\partial F}{\partial z}=\lim_{\Delta z\to 0}\dfrac{e^{A(z+\Delta z)}-e^{Az}}{\Delta z}=e^{At}\lim_{\substack{\Delta t\to 0\\ \Delta z\to 0}}\dfrac{e^{\Delta t}-1}{\Delta z}=e^{At}\lim_{\substack{\Delta t\to 0\\ \Delta z\to 0}}\dfrac{e^{\Delta t}-1}{\Delta t}\dfrac{\Delta t}{\Delta z}=Ae^{At}\sqrt{\varepsilon_0^o\mu_0^o}e^{At}=A\sqrt{\varepsilon_0^o\mu_0^o}e^{2At}\end{cases}, \quad (S23)$$

in which, $\Delta t/\Delta z$ is the reciprocal of the lightspeed, which has been given in Eq. (S22). Collectively, the Maxwell equations can be given as:

$$\begin{cases} \nabla \times \mathbf{H} = \frac{\partial \mathbf{D}}{\partial t} = \varepsilon_0^o F \frac{\partial \mathbf{E}}{\partial t} + A\varepsilon_0^o F \mathbf{E} \\ \nabla \times \mathbf{E} = -\frac{\partial \mathbf{B}}{\partial t} = -\left( \mu_0^o F \frac{\partial \mathbf{H}}{\partial t} + A\mu_0^o F \mathbf{H} \right) \\ \nabla \cdot \mathbf{B} = 0 \\ \nabla \cdot \mathbf{D} = 0 \\ \textbf{Constitutive relation:} \mathbf{D} = \varepsilon_0(t,z)\mathbf{E} \quad \mathbf{B} = \mu_0(t,z)\mathbf{H} \\ \quad \varepsilon_0 = \varepsilon_0^o F \quad \mu_0 = \mu_0^o F \\ \quad F = e^{At} \quad \frac{\partial F}{\partial t} = AF \quad \frac{\partial F}{\partial z} = A\sqrt{\varepsilon_0^o \mu_0^o} F^2 \end{cases} \quad . \quad (S24)$$

Based on the physical analysis, the propagation equations can be given as:

$$\begin{cases} \mathbf{E} = \mathbf{e_x} E_0 e^{-j\left(\omega\sqrt{\varepsilon_0^o \mu_0^o}F\right)z - (A+j\omega)t} \\ \mathbf{H} = \mathbf{e_y} H_0 e^{-j\left(\omega\sqrt{\varepsilon_0^o \mu_0^o}F\right)z - (A+j\omega)t} \end{cases} \quad . \quad (S25)$$

Plug Eq. (S25) it into the curl equations of Eq. (S24), that:

$$\begin{cases} \nabla \times \mathbf{H} = \mathbf{e_y} \varepsilon_0^o E_0 e^{-j\left(\omega\sqrt{\varepsilon_0^o \mu_0^o}F\right)z - (A+j\omega)t} \left[ j\omega F - jA\omega\sqrt{\varepsilon_0^o \mu_0^o}F^2 z \right] = \mathbf{E}\frac{\partial \varepsilon_0}{\partial t} + \varepsilon_0 \frac{\partial \mathbf{E}}{\partial t} \\ \nabla \times \mathbf{E} = \mathbf{e_y} \mu_0^o H_0 e^{-j\left(\omega\sqrt{\varepsilon_0^o \mu_0^o}F\right)z - (A+j\omega)t} \left[ j\omega F - jA\omega\sqrt{\varepsilon_0^o \mu_0^o}F^2 z \right] = -\mathbf{H}\frac{\partial \mu_0}{\partial t} - \mu_0 \frac{\partial \mathbf{H}}{\partial t} \end{cases} \quad . \quad (S26)$$

Eq. (S26) demonstrates that, in time-varied P&P case, the physical results in Eq. (S25) are coincident with the mathematical results on the Maxwell equations in (S24).

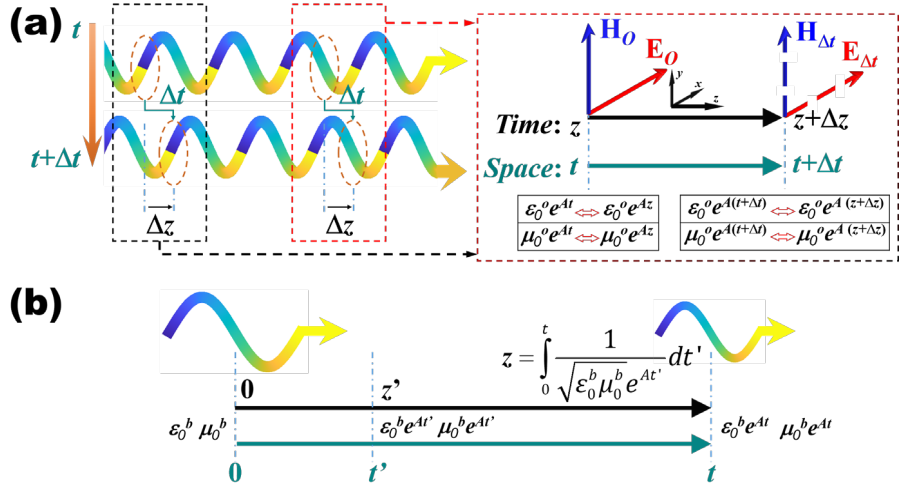

Fig. S2. Propagation of covariant electromagnetic wave. (a) EM waves based on general definition. The wave state is shift with time and position, correspondingly, the P&P for regularizing the EM states is also dependent on time and position. (b) Propagation of single EM wave. An EM wave is generated at $t'=0$ time, and transmits along $z$ direction. The wave can be presented by introducing a unary function, since the position $z$ is dependent on time $t$.

As shown in Fig. S2 (b), the propagation feature of a single EM wave is investigated. The wave is generated at $t'=0$ time and $z'=0$ position, with permittivity and permeability equaling $\varepsilon_0^b$ and $\mu_0^b$, respectively. A factor $F_s$ is introduced to indicate the P&P at time $t'$, that the permittivity and permeability equal to $\varepsilon_0^b F_s(t')$ and $\mu_0^b F_s(t')$, respectively. Obviously, $F_s(t')$ equals to $\exp(At')$. Meanwhile, $F_s(t')$ can also be denoted by $F_s(z')$, in which $z'$ is the distance from the generation position. Based on Eq. (S22), the wave velocity at $t'$ equals to $\exp(At')/(\varepsilon_0^o \mu_0^o)^{1/2}$. For distinguishing the integrable and integral function, a time variable $t$ and a position variable $z$ are introduced. The wave transmits a $z$ distance with a $t$ time. So that the $z$-position equal to the definite integral of $\exp(At')/(\varepsilon_0^o \mu_0^o)^{1/2}$ from $t'=0$ to $t'=t$. Therefore, the propagation feature can be presented by the equations as:

$$\begin{cases} \mathbf{E} = \mathbf{e_x} E_0^b e^{-j\left(\omega\sqrt{\varepsilon_0^b \mu_0^b} F_s\right)z - (A+j\omega)t} \\ \mathbf{H} = \mathbf{e_y} H_0^b e^{-j\left(\omega\sqrt{\varepsilon_0^b \mu_0^b} F_s\right)z - (A+j\omega)t} \\ F_s = e^{At} \\ z = \int_0^t \frac{1}{\sqrt{\varepsilon_0^b \mu_0^b} e^{At'}} dt' \end{cases} \quad (S27)$$

Note that, $F_s$ is a unary function with variable of $t(z)$ or $z(t)$, since $t$ is dependent on $z$. This is different from the function $F$ in Eqs. (S24) and (S25).

**S5: Static electric and magnetic force**

The Maxwell equations of static EM fields are determined based on the electric and magnetic force analysis. Herein, the static EM force is investigated, which is focused on its dependence on (i) time and (ii) position. Therefore, we construct an ideal system with two charges (or currents) with different distance. In addition, the exponential term is taken for the force calculation, in order to that the multiplication calculation can be simplified into the addition operation. The definitions: electric charge $q^* = q^{*o}\exp(S(q^*))$, effect current $\mathbf{J}^*dl = \mathbf{J}^{*o}dl\exp(S(\mathbf{J}^*dl))$, permittivity $\varepsilon_0 = \varepsilon_0^o\exp(S(\varepsilon_0))$, permeability $\mu_0 = \mu_0^o\exp(S(\mu_0))$, electric field $\mathbf{E} = \mathbf{e_E} E^o \exp(S(\mathbf{E}))$, magnetic induction intensity $\mathbf{B} = \mathbf{e_B} B^o \exp(S(\mathbf{B}))$, in which the superscript $o$ denotes the original-value.

We first provide the relevant equations including the static electric and magnetic force:

$$\begin{cases} \mathbf{F}_{ab}^E = \frac{1}{4\pi\varepsilon_0} \frac{q_a^* q_b^*}{r^2} \mathbf{e_r} \\ \mathbf{F}_{ab}^M = \frac{\mu_0}{4\pi} \frac{\mathbf{J}_a^* dl \times (\mathbf{J}_b^* dl \times \mathbf{e_r})}{r^2} \end{cases}, \quad (S28)$$

in which, $\mathbf{F}_{ab}^E$ denotes the static electric force of charge $q^*_b$ from charge $q^*_a$, and $\mathbf{F}_{ab}^M$ denotes the static magnetic force of effect current $\mathbf{J}^*_b dl$ from current $\mathbf{J}^*_a dl$. $\mathbf{J}^* = \tau_0(dq^*/dt)$. $\mathbf{e_r}$ and $r$ are the unit-vector and distance from source $a$ to $b$. The static electric force is analyzed first, as shown in Fig. S3 (a). There exist three charges $q^*_1$, $q^*_2$ and $q^*_0$, in which the force of $q^*_1$ and $q^*_2$ from $q^*_0$ is analyzed only since $\mathbf{F}_{ab} = \mathbf{F}_{ba}$. The distance between $q^*_1$ and $q^*_0$ equals to $z_1$, while that between $q^*_2$ and $q^*_0$ equals to $z_2$. Based on the assumption V, the propagation velocity of static EM fields is equivalent to the lightspeed which has been provided in Eq. (S22), so that $z$ equals to:

$$z = \int_{t-T}^{t} \frac{1}{\sqrt{\varepsilon_0^o \mu_0^o e^{At'}}} dt', \qquad (S29)$$

in which, $t$ is the investigation time with arbitrary value, and $T$ is the time-length of that the electric field propagates from $q^*_0$ to $q^*_1$ or $q^*_2$. Eq. (S29) presents that its propagation velocity also has an exponential term written as $\exp(-At')$, which varies synchronously as other investigated parameters. The values of all $S$ are herein given based on the values at the time $t$, since the $\mathbf{F}_{01}^E$ and $\mathbf{F}_{02}^E$ should be discussed at uniform time. For the analysis of $\mathbf{F}_{01}^E$, the field is generated at $t-T_1$ time. At this time, $S(\varepsilon_0)=A(t-T_1)$, $S(q^*_0)=0$, so that $S(\mathbf{E}_{0\text{-}1})=A(T_1-t)$; meanwhile $S(q^*_1)=0$. During the propagation, with the increase of $S(\varepsilon_0)$, $S(\mathbf{E}_{0\text{-}1})$ decreases from $A(T_1-t)$ to $-At$, while $S(q^*_0)$ and $S(q^*_1)$ are kept unchanged. The field arrives $q^*_1$ at time $t$ at last, of which the $S(\mathbf{E}_{0\text{-}1})=-At$ and $S(q^*_0)=S(q^*_1)=0$. Summarily, $\mathbf{F}_{01}^E=\exp(-At)$. Similar analysis for $\mathbf{F}_{02}^E$, which achieves the same result that $\mathbf{F}_{02}^E=\exp(-At)$. This because that: (i) the field $\mathbf{E}_{0\text{-}2}$ is generated at time $t-T_2$, which has a higher intensity, (ii) the $S(\mathbf{E})$ is linearly decreased that $S(\mathbf{E}_{0\text{-}1})$ and $S(\mathbf{E}_{0\text{-}2})$ has the same gradient to $T$, and (iii) the propagation velocity of static electric field is equal to the lightspeed, which is proportional to $\exp(-At)$. The results above demonstrate that the forces between charges with different distances both equal to $\exp(-At)$, i.e., (i) the static electric force is independent on $T$, i.e., it is independent on position; (ii) the force is decreased with time following $\exp(-At)$. Therefore, the equations of static electric fields can be written as:

$$\begin{cases} \oint_c \mathbf{E} \cdot d\mathbf{l} = 0 \\ \oiint_S \mathbf{D} \cdot d\mathbf{s} = \oiiint_{Vs} \rho dv \\ \textbf{Constitutive relation:} \mathbf{D} = \varepsilon_0 \mathbf{E} \quad \varepsilon_0 = \varepsilon_0^o e^{At} \end{cases}, \qquad (S30)$$

in which, $c$ is a 3D closed curve, $Vs$ is a 3D space with boundary surface $S$, as shown in Fig. 4 (a) in the manuscript.

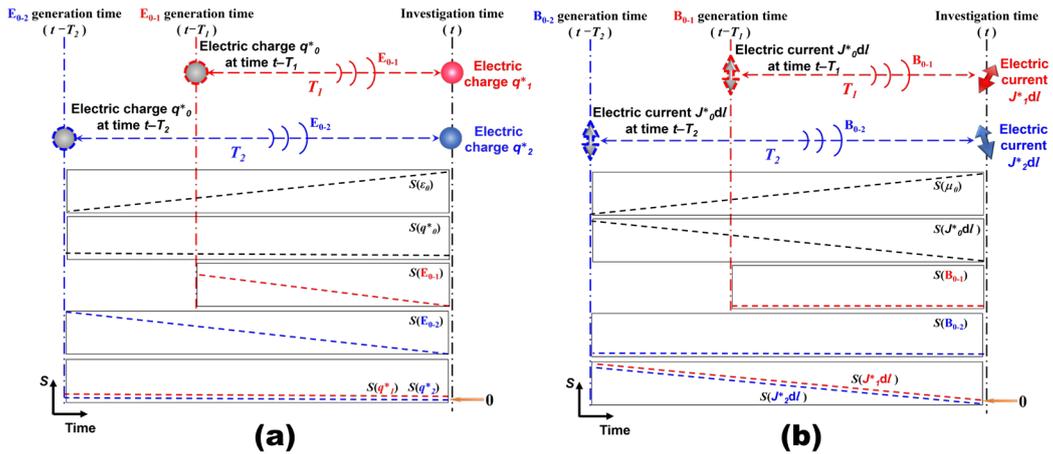

Fig. S3. The static electromagnetic force between charges and currents. (a) Electric force between charges with different distances. (b) Magnetic force between currents with different distances. The static EM force is dependent on time only, therefore, the P&P of static EM field depend on time only.

The static magnetic force is analyzed then, as shown in Fig. S3 (b). There exist three infinitesimal currents $\mathbf{J}^*_1 dl$, $\mathbf{J}^*_2 dl$ and $\mathbf{J}^*_0 dl$, of which the distance between $\mathbf{J}^*_1 dl$ and $\mathbf{J}^*_0 dl$ equals to $z_1$, while that between $J^*_2 dl$ and $J^*_0 dl$ equals to $z_2$. For the analysis of $\mathbf{F}_{01}{}^M$, the field is generated at $t-T_1$ time. At this time $S(\mu_0)=A(t-T_1)$, $S(J^*_0 dl)=A(-t+T_1)$ since $S(\tau_0)=A(-t+T_1)$, so that $S(\mathbf{B}_{0-1})=0$; meanwhile $S(J^*_1 dl)=A(-t+T_1)$. During the propagation, with the increase of $S(\mu_0)$ and decrease of $S(\tau_0)$, $S(\mathbf{B}_{0-1})$ is kept equaling zero, while $S(J^*_0 dl)$ and $S(J^*_1 dl)$ decreases from $A(-t+T_1)$ to $-At$. Note that, though the field is generated by the current $J^*_0 dl$ at $t-T_1$, the $J^*_0 dl$ value for calculating the filed effect should decrease with time. This is because: mathematically, $\tau_0$ is decreased with time; physically, there exist an equivalent relationship between **H** field and the induction current, where **H** field is decreased with time. Based on the EM uniqueness theorem, the induction current is also equivalent to the generation current, so that $J^*_0 dl$ decreases as well. The field arrives $J^*_1 dl$ at time $t$ at last, of which the $S(\mathbf{B}_{0-1})=0$, $S(\mu_0)=At$ and $S(J^*_0 dl)=S(J^*_1 dl)=-At$. Summarily, $\mathbf{F}_{01}{}^M=\exp(-At)$. Similar analysis for $\mathbf{F}_{02}{}^M$, which achieves the same result that $\mathbf{F}_{02}{}^M=\exp(-At)$, because that (i) the currents at the field generation time are higher, (ii) $S(J^*_0 dl)+S(J^*_1 dl)+S(\mu_0)$ and $S(J^*_0 dl)+S(J^*_2 dl)+S(\mu_0)$ has the same gradient to $T$, and (iii) the propagation velocity of static magnetic field is equal to the lightspeed, which is proportional to $\exp(-At)$. The results above demonstrate that the forces between currents with different distances both equal to $\exp(-At)$, i.e., (i) the static magnetic force is independent on $T$, i.e., it is independent on position; (ii) the force is decreased with time following $\exp(-At)$. Therefore, the equations of static magnetic fields can be written as:

$$\begin{cases} \oint_c \mathbf{H} \cdot d\mathbf{l} = \iint_{Sc} \mathbf{J} \cdot d\mathbf{s} \\ \oiint_S \mathbf{B} \cdot d\mathbf{s} = 0 \\ \oiint_S \mathbf{J}_\rho \cdot d\mathbf{s} = -\frac{\partial}{\partial t} \oiiint_{Vs} \rho dv \\ \textbf{Constitutive relation:} \mathbf{B} = \mu_0 \mathbf{H} \quad \mathbf{J} = \tau_0 \mathbf{J}_\rho \quad \mu_0 = \mu_0^o e^{At} \quad \tau_0 = \tau_0^o e^{-At} \end{cases}, \quad (S31)$$

in which, $Sc$ is a 3D surface with its boundary curve $c$, $Vs$ is a 3D space with boundary surface $S$, as shown in Fig. 4 (b) in the manuscript.

The $t$ and $T$ are valued arbitrarily, so that the analysis above includes the general cases. The reason of the independence-on-position: the static EM fields do not have the phase variation, and the magnitude variation of force has been included in the real exponential-term of static EM fields, so that, the P&P of static EM fields are dependent on time only.

## S6: Electromagnetic momentum

Based on the momentum conservation law, the sum of the EM momentum and the mechanical momentum are conservative, which can be presented as:

$$-\frac{d(\mathbf{g}_{em})}{dt} = \frac{d(\mathbf{g}_m)}{dt} = \mathbf{f}, \quad (S32)$$

in which, $\mathbf{g}_{em}$, $\mathbf{g}_m$ and $\mathbf{f}$ denote the EM momentum, mechanical momentum and EM force, respectively. The EM force which can induce the mechanical momentum is written as:

$$\mathbf{f} = \rho \mathbf{E} + \mathbf{J} \times \mathbf{B}. \quad (S33)$$

Plug $\rho = \nabla \cdot \mathbf{D}$ and $\mathbf{J} = \nabla \times \mathbf{H} - \partial \mathbf{D}/\partial t$ into Eq. (S33), and add $(1/\mu_0)(\nabla \cdot \mathbf{B})\mathbf{B}+[\nabla \times (\mathbf{D}/\varepsilon_0)+\partial \mathbf{B}/\partial t] \times \mathbf{D}$ which values zero, achieving that:

$$\mathbf{f} = \left[ \mathbf{B} \times \frac{\partial \mathbf{D}}{\partial t} + \frac{\partial \mathbf{B}}{\partial t} \times \mathbf{D} \right] + \left[ \frac{1}{\varepsilon_0} (\nabla \cdot \mathbf{D}) \mathbf{D} + \nabla \times \left( \frac{\mathbf{D}}{\varepsilon_0} \right) \times \mathbf{D} \right] + \left[ \frac{1}{\mu_0} (\nabla \cdot \mathbf{B}) \mathbf{B} + \nabla \times \left( \frac{\mathbf{B}}{\mu_0} \right) \times \mathbf{B} \right]. \quad (S34)$$

Herein, first, though there does not exist $\rho$ and $\mathbf{J}$ in the equations of covariant EM in open space, they can be used as the force equivalent for the momentum calculation; second, the EM momentum can be investigated in each confined space-region, of which the equations have been provided in Eq. (11) in the manuscript which included $\rho$, $\mathbf{J}$ and the covariant EM fields. We define the terms on the right of Eq. (S34) as $\mathbf{u}$, $\mathbf{r_1}$ and $\mathbf{r_2}$, respectively, from the left to right. The first right term $\mathbf{u}$ equals to:

$$\mathbf{u} = -\frac{\partial (\mathbf{D} \times \mathbf{B})}{\partial t}. \quad (S35)$$

The second term $\mathbf{r_1}$ equals to:

$$\mathbf{r_1} = \frac{1}{\varepsilon_0} \left[ (\nabla \cdot \mathbf{D}) \mathbf{D} + (\mathbf{D} \cdot \nabla) \mathbf{D} \right] - \frac{1}{2\varepsilon_0} \nabla (\mathbf{D} \cdot \mathbf{D}) - (\mathbf{D} \cdot \mathbf{D}) \nabla \left( \frac{1}{\varepsilon_0} \right). \quad (S36)$$

The first term of Eq. (S36) equals to:

$$\frac{1}{\varepsilon_0} \left[ (\nabla \cdot \mathbf{D}) \mathbf{D} + (\mathbf{D} \cdot \nabla) \mathbf{D} \right] = \nabla \cdot \left( \frac{1}{\varepsilon_0} \mathbf{D} \mathbf{D} \right), \quad (S37)$$

while the second term of Eq. (S36) equals to:

$$-\frac{1}{2\varepsilon_0} \nabla (\mathbf{D} \cdot \mathbf{D}) - (\mathbf{D} \cdot \mathbf{D}) \nabla \left( \frac{1}{\varepsilon_0} \right) = -\nabla \cdot \left( \frac{1}{2\varepsilon_0} D^2 \bar{\bar{\mathbf{I}}} \right) - (\mathbf{D} \cdot \mathbf{D}) \nabla \left( \frac{1}{2\varepsilon_0} \right), \quad (S38)$$

in which, $\mathbf{I}$ is the unit tensor. $\nabla(1/\varepsilon_0) \cdot \mathbf{D} = 0$ is used in the derivations of Eqs. (S36) and (S37), since $\nabla(1/\varepsilon_0)$ is perpendicular to $\mathbf{D}$. So that, $\mathbf{r_1}$ can be written as:

$$\begin{cases} \mathbf{r_1} = -\nabla \cdot \left( \frac{1}{2\varepsilon_0} D^2 \bar{\bar{\mathbf{I}}} - \frac{1}{\varepsilon_0} \mathbf{DD} \right) - \nabla \cdot \bar{\bar{\mathbf{\Psi}}}_E \\ \nabla \cdot \bar{\bar{\mathbf{\Psi}}}_E = (\mathbf{D} \cdot \mathbf{D}) \nabla \left( \frac{1}{2\varepsilon_0} \right) \end{cases}. \quad (S39)$$

Similarly, $\mathbf{r_2}$ can be written as:

$$\begin{cases} \mathbf{r_2} = -\nabla \cdot \left( \frac{1}{2\mu_0} B^2 \bar{\bar{\mathbf{I}}} - \frac{1}{\mu_0} \mathbf{BB} \right) - \nabla \cdot \bar{\bar{\mathbf{\Psi}}}_M \\ \nabla \cdot \bar{\bar{\mathbf{\Psi}}}_M = (\mathbf{B} \cdot \mathbf{B}) \nabla \left( \frac{1}{2\mu_0} \right) \end{cases}. \quad (S40)$$

$\mathbf{\Psi}_E$ and $\mathbf{\Psi}_M$ are two momentum-flow-density forms, of which the divergence is a form of EM force. Collectively, from Eqs. (S32), (S34), (S35), (S39) and (S40), we can get:

$$\frac{\partial \mathbf{g}_{em}}{\partial t} = \frac{\partial (\mathbf{D} \times \mathbf{B})}{\partial t} = -\nabla \cdot \left( \frac{1}{2\varepsilon_0} D^2 \bar{\bar{\mathbf{I}}} + \frac{1}{2\mu_0} B^2 \bar{\bar{\mathbf{I}}} - \frac{1}{\varepsilon_0} \mathbf{DD} - \frac{1}{\mu_0} \mathbf{BB} \right) - \nabla \cdot \left( \bar{\bar{\mathbf{\Psi}}}_E + \bar{\bar{\mathbf{\Psi}}}_M \right). \quad (S41)$$

Therefore, $\mathbf{B} \times \mathbf{D}$ is also the momentum density. The first term on the right is the divergence of the classical momentum-flow-density tensor, and the second term is a new form caused by propagation feature of covariant EM wave in time-varied P&P case. In free space, since **E** and **H** decrease following exp(-At), whereas $\varepsilon_0$ and $\mu_0$ following exp(At), so that the $\partial(\mathbf{B}\times\mathbf{D})/\partial t$ is imaginary, i.e., the momentum is time-independent.

**S7: Defined permittivity and permeability based on observation**

Based on the measurements and the International System of Units (IS), the values of $\varepsilon_0$, $\mu_0$ and $\tau_0$ can be defined, of which the observation values are denoted as $\varepsilon_0^{ob}$, $\mu_0^{ob}$ and $\tau_0^{ob}$, respectively. As shown in Fig. S4, the effective-currents $\mathbf{J}^*$ is first defined based on the standards of IS. $\mathbf{J}^*$ with one Ampere is defined as the current intensity when the force between two linear currents, with distance of one-meter, equals to $2\times10^{-7}$ Newton, in which the current length equals to one-meter as well. Meanwhile, based on Eq. (S27), the value of $\mu_0^{ob}$ can be obtained, which equals to $4\pi\times10^{-7}$ NA$^{-2}$. In our general definition system, such a current is regarded as charge-motion, so that $\mathbf{J}^*$ equal to $\mathbf{J}_\rho^*$ based on IS. In this case, $\tau_0^{ob}$ which depends on the measurements at our time equals to one, that $\tau_0^{ob}$ has been actually included into the value of $\mu_0^{ob}$. Third, based on the definition of charge, the $q^*$ with one Coulomb is defined as the charge intensity that induces a one-Ampere current to flow for one second. So that, the value of $q^*$ can be defined. At last, based on Eq. (S27) the value of $\varepsilon_0^{ob}$ is achieved by a measurement of the force between two one-Coulomb charges with distance of one meter. Collectively, all the values of $\varepsilon_0^{ob}$, $\mu_0^{ob}$ and $\tau_0^{ob}$ are defined, i.e., the original values of the constitutive relation are defined based on the measurements at our time.

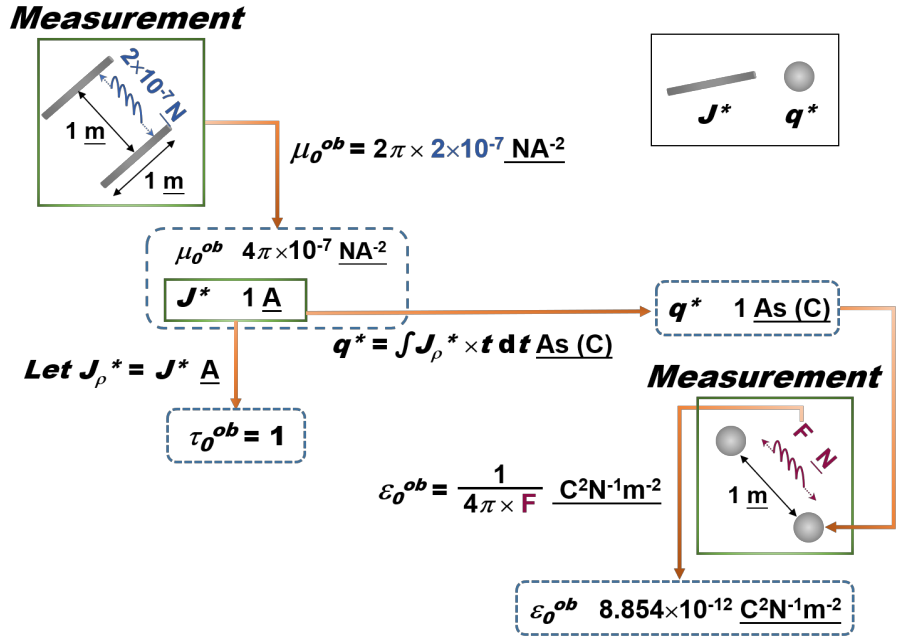

Fig. S4. Flow diagram for defining constitutive relation based on measurements at our time. $\mathbf{J}^*$ and $\mu_0^{ob}$ are defined based on the general definitions of SI. $\tau_0^{ob}$ in our definition system equal to one, since $\mathbf{J}^*$ is regarded as $\mathbf{J}_\rho^*$. $\varepsilon_0^{ob}$ defined by the force measurement between two charges.